%
\let\includefigures=\iffalse
%
\let\useblackboard=\iftrue
%
%
\newfam\black
\input harvmac.tex
\input rotate
\input epsf
\input xyv2
\noblackbox
\includefigures
\message{If you do not have epsf.tex (to include figures),}
\message{change the option at the top of the tex file.}
\def\figin{\epsfcheck\figin}\def\figins{\epsfcheck\figins}
\def\epsfcheck{\ifx\epsfbox\UnDeFiNeD
\message{(NO epsf.tex, FIGURES WILL BE IGNORED)}
\gdef\figin##1{\vskip2in}\gdef\figins##1{\hskip.5in}
\else\message{(FIGURES WILL BE INCLUDED)}%
\gdef\figin##1{##1}\gdef\figins##1{##1}\fi}
\def\DefWarn#1{}
\def\figinsert{\goodbreak\midinsert}
\def\ifig#1#2#3{\DefWarn#1\xdef#1{fig.~\the\figno}
\writedef{#1\leftbracket fig.\noexpand~\the\figno}%
\figinsert\figin{\centerline{#3}}\medskip\centerline{\vbox{\baselineskip12pt
\advance\hsize by -1truein\noindent\footnotefont{\bf Fig.~\the\figno:} #2}}
\bigskip\endinsert\global\advance\figno by1}
\else
\def\ifig#1#2#3{\xdef#1{fig.~\the\figno}
\writedef{#1\leftbracket fig.\noexpand~\the\figno}%
\global\advance\figno by1}
\fi
\useblackboard
\message{If you do not have msbm (blackboard bold) fonts,}
\message{change the option at the top of the tex file.}
\font\blackboard=msbm10
\font\blackboards=msbm7
\font\blackboardss=msbm5
\textfont\black=\blackboard
\scriptfont\black=\blackboards
\scriptscriptfont\black=\blackboardss
\def\Bbb#1{{\fam\black\relax#1}}
\else
\def\Bbb{\bf}
\fi
%
\def\yboxit#1#2{\vbox{\hrule height #1 \hbox{\vrule width #1
\vbox{#2}\vrule width #1 }\hrule height #1 }}
\def\fillbox#1{\hbox to #1{\vbox to #1{\vfil}\hfil}}
\def\ybox{{\lower 1.3pt \yboxit{0.4pt}{\fillbox{8pt}}\hskip-0.2pt}}

\def\rightarrowbox#1#2{
  \setbox1=\hbox{\kern#1{${ #2}$}\kern#1}
  \,\vbox{\offinterlineskip\hbox to\wd1{\hfil\copy1\hfil}
    \kern 3pt\hbox to\wd1{\rightarrowfill}}}
\def\ub{{>}}
\def\lb{{<}}
\def\CY#1{CY$_#1$}

\def\TeV{{\rm TeV}}
\def\GeV{{\rm GeV}}

\def\bz{\bar z}
\def\QC{\Bbb{C}}

\def\QZ{\Bbb{Z}}
\def\p{\partial}

\def\half{{1\over 2}}
\def\Tr{{{\rm Tr~ }}}
\def\tr{{\rm tr\ }}

\def\Im{{\rm Im\hskip0.1em}}

\def\vev#1{\langle{#1}\rangle}

\def\CC{{\cal C}}

\def\CL{{\cal L}}
\def\CM{{\cal M}}
\def\CN{{\cal N}}
\def\CO{{\cal O}}

\def\I{I}

\def\II{\relax{I\kern-.10em I}}
\def\IIa{{\II}a}
\def\IIb{{\II}b}
\def\Coh{{\bf Coh}}

\def\IZ{\relax\ifmmode\mathchoice
{\hbox{\cmss Z\kern-.4em Z}}{\hbox{\cmss Z\kern-.4em Z}}
{\lower.9pt\hbox{\cmsss Z\kern-.4em Z}}
{\lower1.2pt\hbox{\cmsss Z\kern-.4em Z}}\else{\cmss Z\kern-.4em
Z}\fi}
\def\IB{\relax{\rm I\kern-.18em B}}
\def\IC{{\relax\hbox{$\inbar\kern-.3em{\rm C}$}}}
\def\ID{\relax{\rm I\kern-.18em D}}
\def\IE{\relax{\rm I\kern-.18em E}}
\def\IF{\relax{\rm I\kern-.18em F}}
\def\IG{\relax\hbox{$\inbar\kern-.3em{\rm G}$}}
\def\IGa{\relax\hbox{${\rm I}\kern-.18em\Gamma$}}
\def\IH{\relax{\rm I\kern-.18em H}}
\def\II{\relax{\rm I\kern-.18em I}}
\def\IK{\relax{\rm I\kern-.18em K}}
\def\IN{\relax{\rm I\kern-.18em N}}
\def\IP{\relax{\rm I\kern-.18em P}}

%
\def\inbar{\,\vrule height1.5ex width.4pt depth0pt}
\def\mod{{\rm\; mod\;}}

\def\p{\partial}

\def\pb{{\bar \p}}

\font\cmss=cmss10 \font\cmsss=cmss10 at 7pt
\def\IR{\relax{\rm I\kern-.18em R}}

\def\Vol{{\rm Vol}}

\def\BZ{\QZ} 
\def\BP{\IP}

\def\BC{\QC}

\def\lp10{l_P^{10}}
\def\lp11{l_P^{11}}
\def\R11{R_{11}}

\newbox\tmpbox\setbox\tmpbox\hbox{\abstractfont RUNHETC-2002-09}
\Title{\vbox{\baselineskip12pt\hbox to\wd\tmpbox{\hss
hep-th/0303194}\hbox{RUNHETC-2003-09}}}
{\vbox{
\centerline{The statistics of string/M theory vacua}}}
\smallskip
\centerline{
Michael R. Douglas}
\smallskip
\bigskip
\medskip
\centerline{Department of Physics and Astronomy, Rutgers University,
 Piscataway, NJ 08855-0849 USA}
\medskip
\centerline{I.H.E.S.\footnote{$^{\dagger}$}{
Louis Michel Professor}, Le Bois-Marie, Bures-sur-Yvette, 91440 France}
\bigskip

\noindent
We discuss systematic approaches to the classification of string/M
theory vacua, and physical questions this might help us resolve.  To
this end, we initiate the study of ensembles of effective Lagrangians,
which can be used to precisely study the predictive power of string
theory, and in simple examples can lead to universality results.
Using these ideas, we outline an approach to estimating the
number of vacua of string/M theory which can realize the Standard Model.

\Date{March 2003}
\def\np{{\it Nucl. Phys.}}

\def\pr{{\it Phys. Rev.}}
\def\pl{{\it Phys. Lett.}}

\def\cmp{{\it Comm. Math. Phys.}}
%
\nref\ach{B. Acharya, M. Aganacic, K. Hori and C. Vafa,
hep-th/0202208.}
\nref\achflux{B. Acharya, hep-th/0212294.}
\nref\acharyanew{B. Acharya, to appear.}
\nref\sagnotti{C. Angelantonj and A. Sagnotti,
Phys. Rept. 371 (2002) 1-150; hep-th/0204089.}
\nref\agm{P.~S. Aspinwall, B.~R. Greene, and D.~R. Morrison,
\np\ {\bf B416} (1994) 414--480.}
\nref\aspdoug{P.~S. Aspinwall and M.~R.~Douglas,
JHEP {\bf 0205} 031 (2002); hep-th/0110071.}
\nref\banks{T. Banks, hep-th/0211160.}
\nref\banksglo{T. Banks and L.J. Dixon,
\np\ {\bf B307}, 93 (1988).}
\nref\bdd{T. Banks, M. Dine and M. R. Douglas, hep-ph/0112059.}
\nref\bdg{T. Banks, M. Dine and M. Graesser, hep-ph/0210256.}
\nref\bdm{T. Banks, M. Dine and L. Motl, 
JHEP 0101 (2001) 031; hep-th/0007026.}
\nref\bea{C.~E.~Beasley and M.~R.~Plesser,
JHEP {\bf 0112}, 001 (2001), hep-th/0109053.} 
\nref\beckerG{K. Becker and M. Becker,
Nucl.Phys. B477 (1996) 155-167; hep-th/9605053.}
\nref\beckers{K. Becker, M. Becker, M. Haack and J. Louis, hep-th/0204254.}
\nref\bd{D. Berenstein and M. R. Douglas, hep-th/0207027.}
\nref\zelditch{P. Bleher, B. Shiffman and S. Zelditch, 
{\it Comm.\ Math.\ Phys.},   208 (2000), 771.}
\nref\bousspol{R. Bousso and J. Polchinski,
JHEP 0006 (2000) 006, hep-th/0004134.}
\nref\brown{J. D. Brown and C. Teitelboim, 
\np\ B297 (1988) 787.}
\nref\cachdual{
F.~Cachazo, B.~Fiol, K.~A.~Intriligator, S.~Katz and C.~Vafa,
\np\ B {\bf 628}, 3 (2002), hep-th/0110028.}
\nref\cdsw{F.~Cachazo, M.~R.~Douglas, N.~Seiberg and E.~Witten,
JHEP 0212 (2002) 071; hep-th/0211170.}
\nref\csw{F.~Cachazo, N.~Seiberg and E.~Witten,
JHEP 0302 (2003) 042; hep-th/0301006.}
\nref\chsw{P.~Candelas, G.~Horowitz, A.~Strominger, and E.~Witten,
\np\ {\bf B258} (1985) 46--74.}
\nref\candelas{P. Candelas, X. C. de la Ossa, P. S. Green, and 
  L. Parkes,  \np\ {\bf B359} (1991) 21.}
\nref\uranga{J. F. G. Cascales and A. M. Uranga, hep-th/0303024.}
\nref\coleman{S. Coleman and F. De Luccia, \pr\ D21 (1980) 3305.}
\nref\cox{D.~A. Cox and S.~Katz,
{\it Mirror Symmetry and Algebraic Geometry}, Mathematical Surveys
  and Monographs~{\bf 68}, AMS, 1999.}
\nref\csu{M. Cvetic, G. Shiu, A.M. Uranga, hep-th/0107166.}
\nref\cvetic{M. Cvetic, private communication.}
\nref\dasgupta{K.  Dasgupta, G. Rajesh and S. Sethi,
JHEP 9908 (1999) 023, hep-th/9908088.}
\nref\denef{F. Denef, JHEP 0008 (2000) 050, hep-th/0005049.}
\nref\ddg{D.-E. Diaconescu, M. R. Douglas and J. Gomis,
hep-th/9712230.}
\nref\dd{D.-E. Diaconescu and M. R. Douglas,
hep-th/0006224.}
\nref\dv{R. Dijkgraaf and C. Vafa, hep-th/0208048.}
\nref\donagi{R.~Y. Donagi,
Asian J. Math. {\bf 1} (1997) 214--223, alg-geom/9702002.}
\nref\donkron{S.K. Donaldson and P.B. Kronheimer, 
{\it The Geometry of Four-Manifolds}, Oxford: Clarendon Press.}
\nref\dfr{M.R. Douglas, B. Fiol and C. R\"omelsberger,
``The spectrum of BPS branes on a noncompact Calabi-Yau,''
hep-th/0003263.}
\nref\trieste{M.R. Douglas, lectures at the 2001 Trieste spring
school, available on the web at 
{\tt http://www.ictp.trieste.it} .}
\nref\houches{M. R. Douglas, 
``String Compactification with $N=1$ Supersymmetry,''
in C. Bachas, A. Bilal, M. R. Douglas and N. A. Nekrasov, eds.,
{\it Unity from Duality: Gravity, Gauge Theory and Strings,}
Les Houches 2001, North Holland.  Not available on the web.}
\nref\jhstalk{M. R. Douglas, Lecture at JHS60, October 2001, Caltech.
Available on the web at {\tt http://theory.caltech.edu}.}
\nref\stringscam{M. R. Douglas, Lecture at Strings 2002, Cambridge UK.
Available on the web at {\tt http://strings.cam.ac.uk}.}
\nref\dougashok{M. R. Douglas, S. Ashok, and others, work in progress.}
\nref\dgjt{M.~R.~Douglas, S.~Govindarajan, T.~Jayaraman, and A.~Tomasiello,
hep-th/0203173, to appear in \cmp.}
\nref\dm{M.~R.~Douglas and G.~Moore, hep-th/9603167.}
\nref\dougzel{M. R. Douglas, B. Shiffman and S. Zelditch, work in progress.}
\nref\dougwit{M. R. Douglas, E. Witten and others, work in progress.}
\nref\dougzhou{M. R. Douglas and C.-G. Zhou, to appear.}
\nref\fmw{R.~Friedman, J.~W. Morgan, and E.~Witten,
J. Alg. Geom. {\bf 8} (1999) 279--401, alg-geom/9709029.}
\nref\fulton{W. Fulton, {\it Intersection Theory}, Springer, 1998.}
\nref\feng{J. L. Feng, J. March-Russell, S. Sethi and F. Wilczek,
\np\ B602 (2001) 307; hep-th/0005276.}
\nref\fengdual{B.~Feng, A.~Hanany, Y.~H.~He and A.~M.~Uranga,
JHEP {\bf 0112}, 035 (2001), hep-th/0109063.}
\nref\fiol{B .~Fiol, JHEP 0207 (2002) 058; hep-th/0205155.}
\nref\friedmann{T. Friedmann and E. Witten, hep-th/0211269.}
\nref\gepner{D. Gepner, \np\ B296 (1988) 757.}
\nref\gkp{S. B. Giddings, S. Kachru and J. Polchinski,
\pr\ D66 (2002) 106006; hep-th/0105097.}
\nref\gimpol{E. Gimon and J. Polchinski,
\pr\ D54 (1996) 1667; hep-th/9601038.}
\nref\gv{R. Gopakumar and C. Vafa,
Adv. Math. Theor. Phys. 3 (1999) 1415; hep-th/9811131.}
\nref\gsw{M.B. Green, J.H. Schwarz, E. Witten,
{\it Superstring Theory}, Volume II. Cambridge: Cambridge
University Press.}
\nref\greenschwarz{M.~B. Green and J.~H. Schwarz, \pl\ 149B, 117.}
\nref\greene{B.~R. Greene,
``String Theory on Calabi--Yau Manifolds,''
in C.~Esthimiou and B.~Greene, editors, {\it Fields, Strings and
  Duality, TASI 1996}, pages 543--726, World Scientific, 1997, 
hep-th/9702155.}
\nref\markgross{M. Gross, private communication.}
\nref\gvw{S. Gukov, C. Vafa and E. Witten,
\np\ B584 (2000) 69; hep-th/9906070.}
\nref\hornemoore{J. H. Horne and G. Moore, 
\np\ B432 (1994) 109; hep-th/9403058.}
\nref\ibanez{L. E. Ibanez, hep-ph/0109082.}
\nref\johnson{C. Johnson, hep-th/0007170.}
\nref\kst{S. Kachru, M. Schultz and S. Trivedi, hep-th/0201028.}
\nref\linde{S. Kachru, R. Kallosh, A. Linde and S. Trivedi, hep-th/0301240.}
\nref\threshold{V. Kaplunovsky, hep-th/9205070.}
\nref\katzcom{S. Katz, private communication.}
\nref\king{A.~D.~King, Quart. J. Math. Oxford (2), 45 (1994), 515-530.}
\nref\kobayashi{S. Kobayashi, {\it Hyperbolic Complex Spaces},
Springer 1998.}
\nref\kreusch{M. Kreuzer and H. Skarke, 
Adv. Theor. Math. Phys. 4 (2002) 1209; hep-th/0002240.}
\nref\malda{J. Maldacena, Adv. Theor. Math. Phys. 2 (1998) 231;
hep-th/9711200.}
\nref\moore{G. Moore, hep-th/9807087.}
\nref\ov{H. Ooguri and C. Vafa, 
\np\ B641 (2002) 3; hep-th/0205297.}
\nref\roemel{C. R\"omelsberger, hep-th/0111086.}
\nref\psflux{J. Polchinski and A. Strominger,
\pl\ B388 (1996) 736-742, hep-th/9510227.}
\nref\polchinski{J. Polchinski, ``TASI Lectures on D-Branes,''
hep-th/9611050.}
\nref\ps{J. Polchinski and M. J. Strassler, hep-th/0003136.}
\nref\moduliproblem{J. Polonyi, as referenced in \bdg.}
\nref\schimmrigk{R. Schimmrigk, alg-geom/9612012.}
\nref\sei{N. Seiberg, \np\ B {\bf 435}, 129 (1995);
hep-th/9411149.}
\nref\ss{Y. Shadmi and Y. Shirman,
Rev. Mod. Phys. 72 (2000) 25-64.}
\nref\stromtor{A. Strominger,
\np\ B274 (1986) 253.}
\nref\sv{A. Strominger and C. Vafa, \pl\ B379 (1996) 99;
hep-th/9601029.}
\nref\spinglass{F. Tanaka and S. F. Edwards,
J. Phys. F: Metal Phys. 10 (1980) 2769.}
\nref\trivedi{P. K. Tripathy and S. P. Trivedi, hep-th/0301139.}
\nref\vafabh{C. Vafa, Adv. Theor. Math. Phys. 2 (1998) 207; hep-th/9711067.}
\nref\vafadual{C. Vafa, J. Math. Phys. 42 (2001) 2798; hep-th/0008142.}
\nref\wessbagger{J. Wess and J. Bagger,
{\it Supersymmetry and Supergravity}, 
Princeton Univ. Press., 1990.}
\nref\wittenindex{E. Witten, \np\ B202, 253, 1982.}
\nref\witteninst{E. Witten, \np\ B460 (1996) 541; hep-th/9511030.}
\nref\zelreview{S. Zelditch, math.CA/0208104.}
%
%
\newsec{General introduction}

String/M theory, the unified web of dual theories which subsumes
superstring theory and supergravity, is by far the best candidate we
have for a unified theory of fundamental physics.  It describes
quantum gravity, and making very simple compactifications, can
lead to supersymmetric grand unified theories remarkably close to
those postulated as natural extensions of the Standard Model which
solve the hierarchy problem \gsw.

This agreement, while impressive, is still only qualitative.  More
precise comparison has foundered on the ``vacuum selection problem.''
Despite the unity of the theory, string/M theory appears to describe a
very large number of four dimensional (and other) vacua with
inequivalent physics, most of which clearly do not describe our universe.
At present we have no clue which one is relevant, or how to find it.

There are different points of view about how this problem will be
solved.  The simplest is to say that if we search through the
possibilities, we will eventually find the right vacuum, meaning the
one which agrees with our data, and we can then ignore the others.
This point of view is admirably direct, and to some extent we will
advocate it, but it appears that the number of vacua is so large and
the problem of constructing and testing them is so complicated that
one needs to better organize the problem to have any hope of success.

Many believe that rather than do an exhaustive search, we need to find
a ``Vacuum Selection Principle,'' an {\it a priori} condition which
will tell us which vacuum to consider.  Now at present there are no
good candidates for this principle.  Based on our present
understanding of string/M theory, it appears that the only obvious
candidate principle, nonperturbative consistency, is not very
selective.  One can hope that a selection principle will emerge from
the study of cosmology, but it can just as well be argued that
cosmology will only lead to constraints of the same general type as
those we already employ in phenomenology, namely tests which must be
satisfied by our vacuum or by the effective Lagrangian in some
neighborhood of our vacuum in configuration space,
but which do not give much {\it a priori}
guidance about which vacuum to look at.  Such principles are valuable
but do not cut through the practical difficulties we just cited.

One can clarify many of the issues and obtain a much more well-defined
problem by taking the opposite position, which is that there is no
Vacuum Selection Principle in the sense we just discussed.  Rather,
one must simply enumerate string/M theory vacua and test each one
against all constraints inferred from experiment and observation.
While this task may seem like searching for a needle
in a haystack, this does not make it impossible or uninteresting.
After all, with modern technology (say a harvester equipped with a
magnet) one can find needles in haystacks without much difficulty.  It
may turn out that upon approaching the problem systematically, we will
in fact find easy ways to identify the vacua which might be relevant
for our world.

In my own work, this point of view evolved in
\refs{\jhstalk,\houches}, and led to the idea 
that one must get a good overall
picture of how many vacua the theory has and a statistical description
of their properties, to guide any such search.  The present paper is
an introduction to these ideas and some lines of work which this point
of view has inspired, details of which will appear elsewhere
\refs{\dougzel,\dougashok}.

In talking to my colleagues, I find that this point of view is
sometimes considered to be defeatist, abandoning any hope of ``real
explanation.''  I believe this is not right, and to explain why
I have provided a ``philosophical introduction'' in section 2,
expanding on a discussion in \houches.  Although ``explanation'' is
a subjective concept, the most important question along these lines is
whether string/M theory is falsifiable given sufficient theoretical
understanding and sufficient data.  In fact when one considers this
point carefully, one realizes that it may not be falsifiable.  The
basic problem is that we have not ruled out the possibility that
string/M theory contains a large or even infinite set of vacua which
arbitrarily well approximate the Standard Model and any of its
extensions we might hope to establish experimentally. \jhstalk\ %
While it is reasonable to believe that this is false and that string/M
theory is falsifiable, we intend to argue that this is a question which
can be subject to theoretical analysis and settled, we suspect long
before ``the right vacuum'' is identified.

To explain our point, let us imagine the logically simplest possible
discussion of ``string phenomenology.''  It would be to show that $N$
different vacua of string/M theory lead to Standard Model-like
physics, but with many different values of the couplings, uniformly
distributed in the space of possible couplings (we will make this more
precise in section 5).  Now the basic number characterizing our
observational knowledge of the Standard Model is the volume in
coupling space consistent with observations, measured in natural
units, $O(1)$ for dimensionless couplings and $O(M_{pl}^n)$ for a
coupling of mass dimension $n$.  If we include as couplings the Higgs
mass and the cosmological constant, this number is of order
$10^{-120-40-10-9-9-50} \sim 10^{-238}$, where we count as independent
the probability for a model to realize the observed cosmological
constant, Higgs mass, fine structure constant, electron and proton
mass, and a product of all other Standard Model couplings (being
generous in the assumed accuracy here).  This is a very high
precision, but suppose string/M theory led to $10^{1000}$ vacua which
matched the Standard Model gauge group and low energy spectrum.  If
so, it is likely that, in the absence of a selection principle,
string/M theory would lead to no testable predictions at all.  

Although this number of models may seem absurdly high, of course by
multiplying a modest number of independent choices, one can easily
produce much larger numbers.  Furthermore, our estimate for the
likelihood of matching the Standard Model is far too low, as we did
not even take supersymmetry into account.  In any case, at present we
have no meaningful estimate of the number of vacua which might
approximate the Standard Model.  Indeed, we have no real argument that
the number is finite; and we will argue below that if we are too
inclusive in our definitions, string/M theory probably leads to an
infinite number of vacua.

Thus, the primary question along these lines is to somehow estimate the
number of string/M theory vacua which should approximate the Standard
Model.  This is obviously difficult and we will not claim to have
even properly formulated the question here.  What we will do is make
some first steps towards properly formulating it, and try to make the
case that this goal could be easier than actually finding all the
vacua or even the one which describes our world.  For the reasons we
just discussed, it might even turn out that the answer to this
question will force us to drastically re-evaluate the simple idea that
``we need to find the vacuum which describes our world.''

Having further justified our approach in section 2, we begin by giving
a very sketchy overview of the problem of string/M theory
compactification in section 3.  Although necessarily somewhat
simplistic, such an overview is necessary to give any content to the
subsequent discussion.  We also justify some of our further
assumptions.  First, as has been argued by many, at our present level
of theoretical sophistication our only hope of making statements of
the generality we need is to assume that nature has spontaneously
broken supersymmetry.  For reasons we discuss, we consider models
which arise from type \I\ and type \II\ orientifold compactification
on Calabi-Yau, and develop a picture based on recent work on branes
and vector bundles, and on compactifications with flux.

We will also discuss some basic estimates for numbers of vacua,
which can be applied even in the absence of detailed understanding
of quantum corrections.  In particular, one can get estimates using
facts about the topology of moduli spaces, or using combinatorics
of brane constructions.

An important input into any claim which uses the total number of vacua
to estimate the fraction of the vacua which look like the Standard
Model, is knowledge of how the vacua are distributed among the
possibilities.  We will argue that flux stabilization of moduli typically
leads to ``uniform'' distributions.

Another point which our discussion will make, which we think will be
uncontroversial, is that the problem of constructing and classifying
string/M theory compactifications is very complicated.  Furthermore,
if one tries to summarize it in the language of effective Lagrangians,
one is led to strongly suspect that the quantities which enter
(superpotential and K\"ahler potential) are very complicated
functions.  Now there is a time-honored way to deal with certain types
of complexity in theoretical physics.  It is the statistical approach,
in which we introduce ensembles of randomly chosen systems, and study
expected values of the quantities of interest.  The great advantage is
of course that these ensembles can be far easier to formulate and
study than the true system, while the hope is that some properties of
the true system will hold in the average system, and will be visible
in the expected values.  Sometimes this approach works, and in
favorable cases one even finds that some quantities of interest are
universal, meaning that they do not depend on the details of the
ensemble but only on a few parameters which can be determined.
Clearly having such quantities would be of great value.

In section 4, we formulate some simple ensembles of $\CN=1$ effective
supergravity Lagrangians.  We also pose some questions which might be
interesting to study along these lines, and might show universality.
In fact, there is an ensemble which has already been studied by
mathematicians (for applications to quantum chaos) which can be
adapted to the problem at hand, and this will enable us to actually
cite a universality result of this type, governing the distribution of
supersymmetric vacua.

We regard such ensembles as tools for understanding and steps towards
our primary goals of properly understanding the actual set of string/M
theory vacua, and estimating the number which could describe the real
world.  We suggest an estimate for this number, at least from one
class of construction, in section 5.  Although there will be gaps in
this discussion and we will not claim that our estimate is reliable,
we felt this exercise was useful to illustrate the use of ensembles,
and to get some preliminary sense of the problem and show which parts
of it are better under control and which parts are less so.  Indeed,
we will not rule out the possibility that the number of vacua is large
enough to spoil testability, again in the absence of other selection
principles.

In section 6, we briefly summarize and conclude.

\newsec{A more philosophical introduction}

This section is an expansion on points made in \houches.  It could be
skipped by readers with a distaste for this sort of discussion.

As we mentioned in the introduction, there is a widespread feeling
that a ``theory of everything'' should make unique predictions for
the physics we observe.  String/M theory as we understand it now
does not do this, and it is this lack which is often cited as the
reason why a ``Vacuum Selection Principle'' should exist.  Of course,
this argument in itself is simply wishful thinking.

Let me indulge in a little analogy.  Suppose we were characters in a
1930's science fiction story, who lived on an electron orbiting a
nucleus.  By observation, we might discover that our particular
nucleus had 9 electrons orbiting it.  We might even formulate the
Schr\"odinger equation and find that our atom was a particular
solution.  Even without observing other atoms, by mathematical
analysis of this equation, we would discover the possibility of
hydrogen, helium and so on; the amazing fact would emerge that
(granting the quantization of electric charge, and taking relativistic
effects into account) this equation only had
about 100 solutions of the general type we observed.  

Having gotten that far, we could spend a long time looking for the
``fluorine selection principle'' which completes the story.  Of
course, if the physics were really governed by the Schr\"odinger
equation, we would never find it.

Although the analogy is a bit forced, the kernel of truth in it is
that, according to our present understanding, the consistent
unification of quantum mechanics and gravity through string/M theory
seems to lead to a definite set of solutions which resemble our world.
This is already a great deal of predictivity, and we should see how
far it can take us.

Of course, unique predictivity is not at all required for a theory
to be scientifically testable and falsifiable.  It is far more than
we expect from most theories.  Still, one hopes that a theory with
``no free parameters'' could do better than most theories.

The sense in which string/M theory is better than generic quantum
field theory, relies on the idea that vacua are local minima, or
approximate minima, of the effective potential.  While all coupling
constants must be vacuum expectation values of fields, since a generic
effective potential in a nonsupersymmetric theory will have isolated
minima, all of these expectation values will take definite values in a
given vacuum.  Even if the minimum is not unique, one still obtains
a list of potential predictions, one for each minimum, and the theory
can be falsified.

Now the expectation that a minimum is isolated, and thus that
couplings are stabilized, is a generic statement which could
have exceptions.  More precisely, coupling constants could in
principle be time dependent, but this is unnatural (for the Standard
Model couplings and especially the fine structure constant)
\bdd.  Observation of such a variation would therefore lead to much
stronger constraints on the vacuum than any mere observation of a
particular fixed value, and one will have testability in the same sense.
Similar comments apply to ``dark energy'' or ``quintessence'' (which
is not quite as unnatural).

Of course, fitting observation provides many ``vacuum selection
principles'' in a weaker sense.  The most optimistic scenario is that
future accelerator experiments will provide direct evidence for
string-like or other structure which is not naturally described by
four dimensional field theory, and which will give us information
which will directly constrain the choice of vacuum.  While this is
certainly the most attractive scenario, there are many others,
including the original ones described in \gsw, in which this appears
impossible: the energy scale of the new effects is far beyond any
conceivable experiment.  Given that the only new energy scales we have
evidence for at present are $M_{Planck}\sim 10^{19}\ \GeV$ and
$M_{GUT}\sim 10^{16}\ \GeV$, these latter scenarios must be taken
seriously and are perhaps even preferred.

Even without such direct evidence,
fitting the known data would already be quite constraining.  Besides
the obvious tests of fitting the spectrum and couplings, one can
propose indirect ones, for example to fit our present understanding of
cosmology.  An extreme example of such a test is the idea that the
vacuum energy, literally defined as the value of the effective
potential at its minimum, must in fact reproduce the observed
cosmological constant.  Other examples include the ideas that one must
obtain inflation, or that uncharged scalars in a range of masses
around $1 \TeV$ are not allowed \moduliproblem.\foot{To properly
discuss cosmological tests, one must grant that in early cosmology the
universe does not minimize the effective potential, and in general one
needs more than the low energy effective field theory.  In this
context, when we talk about a ``vacuum,'' we mean not just the
effective field theory at the minimum but whatever computations in the
underlying theory (string/M theory) are required to make the test; of
course the results of such computations will depend drastically on the
choice of minimum one is working near.  A simple and sometimes valid
picture is that one is following some trajectory which ends up at the
minimum or approximate minimum and is seeing the effective field
theory along this trajectory.  Such tests are appropriate within our
``no Vacuum Selection Principle'' assumption, but one should regard a
vacuum as passing the tests if it can do so for any of the initial
conditions within a chosen subset of non-zero measure.}  If a unique
string/M theory vacuun (or none at all) were to pass these tests, the
vacuum selection problem would be solved in a practical sense.

One difference between the principles we actually know about and the
hoped-for ``Vacuum Selection Principle'' is that to our present
understanding, no one of these tests seems more fundamental or key than
the others.  But the biggest practical difference between the two
ideas is that the ``vacuum selection principles'' are {\it a
posteriori} tests which require constructing and studying a vacuum in
great detail.  

With the present state of the art, even the basic construction and
analysis of one or a few vacua is a research program requiring
several man-months of effort to complete and several papers to
describe.  Very few vacua have been studied on even
the level discussed in \gsw.
One can try to develop better techniques, but one should
realize that the standard questions which are addressed by such
analyses, motivated fairly directly by comparison with experiment,
such as the structure of scales and hierarchies, the gauge group and 
charged matter content, indeed require at least a page to answer.
One cannot hope to analyze a model in less time than it takes to
read and understand the results, and the number of models is
such that even this is not possible for each model.

One might at least hope that some of the tests are {\it a priori},
meaning that one can restrict attention from the start to vacua with
the given property.  We are certainly doing this in restricting
attention to models with (according to present experiment) four
dimensions.  Another example is that in most perturbative
constructions, the low energy gauge group is more or less fixed very
early in the analysis, and it is easy to exclude models in which this
is (say) too small to embed the Standard Model.

A more ambitious hope along these lines is to ``engineer'' vacua,
taking observations as encoded in the Standard Model and its
extensions and directly building models which reproduce the
rough features of the observations, expecting string/M theory to
then tell us the fine details, such as values of couplings.

It is hard to argue against these ideas, which are good to the extent
that they can be implemented and actually constrain the problem.  The
main problem with them is that too much of the problem seems
unconstrained by observation: there are many very different ways to
realize the observed matter, and many ``hidden'' sectors which
directly influence the couplings and other data we hope to predict.

We should say a brief word about the anthropic principle here and will
only say that, while interesting, we feel this is raising a different
question than the one we are discussing.  Surely it is true that most
of the possible universes which come out of string/M theory do not
look like four large dimensions, do not lead to macroscopic structure
formation, or do not lead to environments suitable for any sort of
life.  Any of these conditions will also lead to tight constraints on
the physical laws, and it is interesting to explore these.  On the
other hand, we inhabit a large four dimensional universe with specific
physical laws that we determine by observation, laws surely much more
specific than any anthropic consideration we can seriously study will
lead to, and it is not at all clear whether any string/M theory vacuum
reproduces these laws.  If we know the laws, and if string theory has
a precise formulation, then the question of whether string theory can
reproduce the laws or not can in principle be answered without ever
appealing to any of the anthropic considerations.

Having defined our problem, we can say in a nutshell the main new idea
we will introduce in this work: it is to gain insight and results
bearing on the problem, by studying ensembles of theories which
approximate the true ensemble of vacua coming out of string/M theory.
We will discuss various approximation schemes below.  One general
approach is to imagine the true set of vacua as a ``sum of delta
functions in theory space,'' and to approximate these delta functions
with a similar but more general distribution of weights.  We also make
various less systematic approximations.

Although at first the distinction may seem to be splitting hairs, we
stress that the ensembles we are introducing are not probability
distributions but rather describe the distribution of vacua in
particular regions of ``theory space,'' without any implication that
one vacuum is ``more likely'' than another.  Such an ensemble is not
normalized to unit probability; rather each vacuum contributes unit
measure and the total distribution is normalized to the total number
of vacua.

The main reason we emphasize this distinction is the obvious point
that given that we observe one particular vacuum, the ensembles we
will discuss are not directly observable.
Thus certain questions which might seem to be the natural applications
of an ensemble of vacua, do not really make sense.  The prime example
is the question ``what is the probability to realize vacuum X.''  One
might imagine pushing ideas such as the ones we discuss or related
ones in quantum cosmology to the point of making statements such as
``the probability of our vacuum is $10^{-47}$, while the probability
of vacuum Y which we do not live in is $10^{-42}$ (or maybe
$10^{-52}$), but it is not clear to us what scientific conclusions one
can draw from such statements.\foot{This is not to say that a ``wave
function of the universe,'' which might be described by a
complex-valued distribution of the type we will discuss, would be
uninteresting to study.  We are just not convinced that it should be
used to compute relative probabilities of vacua.}  In particular, even
if it turned out that we live in a highly improbable vacuum according
to both quantum cosmological and anthropic considerations based on
string theory, we do not think this could be considered as evidence
against string theory.\foot{This is in the absence of a competing
theory.  Of course, if one finds a competing theory which can also
explain the observations, one must judge which appears more
predictive, plausible, natural, likely, or whatever.  All potential
competitors which we know about have far more arbitrariness than
string/M theory, in particular they depend on pre-specified adjustible
couplng constants, and this is already a reason to suspect they will
be less predictive.  One should reexamine all this if a better
competitor emerges, but it seems useless to try to provide any
guidance for this in advance.}  Only if the vacuum we live in is
literally not a possible prediction of string/M theory, meaning either
that ``it does not appear on the list'' or that we propose a well
motivated assignment of probabilities to vacua which gives it zero
probability, could we consider this as falsifying string/M theory.

Of course, there are other applications of ensembles for which the
probability interpretation makes sense.  For example, because of
statistical uncertainty in experiment and observation, we will never
get us a precise set of laws to try to make contact with, and this
uncertainty could also be summarized in an ensemble of effective
theories.  Thus, the real problem is to see if the subset of theories
which are ``reasonable'' fits to data contains any theories in the
ensemble of string/M theory vacua.

Experimental uncertainty is important, but it should be clear
that we are introducing a new type of uncertainty, more as a
theoretical device, and different from experimental uncertainty.  The
statistics which enters in understanding experimental uncertainty is
of course the bread and butter of experimental analysis and
phenomenology, and a fairly mature subject.  As we did in the
introduction, we need to talk about experimental accuracy to make our
points, and to precisely define ``the accuracy to which we know the
Standard Model'' we would need to formalize this, but we do not need
such a precise definition to make our main points here.

Finally, one idea which is certainly part of this circle of ideas is
what we call the ``purely statistical'' scenario: that all the
observed structure and couplings of the Standard Model emerge as one
undistinguished choice from a completely uniform distribution of low
energy theories.  While in the absence of any other idea this might be
attractive, for example to solve the cosmological constant problem, it
may seem implausible or even repellent when applied to other aspects
of physics which seem to point clearly to structure, such as the
unification of gauge couplings.  This may be, and our job as
physicists is in part to find structure, but it should be
realized that it may simply be that string/M theory contains both
vacua which realize observed physics by mechanism and vacua which
realize it statistically, and this would be important to know.  

We will also argue that the statistical approach is relevant for
theories with mechanisms; one can try to estimate the number of grand
unified models, the number of models with low energy supersymmetry,
and so on.  One can use ensembles to gain information about what
mechanisms are easy to realize in string theory and what mechanisms
are difficult, and whether the features explained by the mechanism
justify the choices involved.  In this sense, one could think of
an ensemble which accurately represented the set of string/M theory
vacua as providing a ``stringy'' idea of naturalness.  The discussion
in section 5 is intended to illustrate this idea.

Although for clarity we assume throughout this work that there
is no ``Vacuum Selection Principle,''\foot{One might call this
claim the ``Ultimate Copernican Principle.''} of course we do not
know whether there is one or not.
Let us conclude by giving what in our opinion are the best
arguments ``for'' and ``against'' a Vacuum Selection Principle.

The best hope is the esthetically motivated hope that observed physics
arises from a particularly ``symmetric'' or ``natural'' string
compactification.
Physicists have clearly been lucky in that the fundamental laws are
comprehensible at all, so why shouldn't our luck hold to the end.

Against this hope, one can make the claim that the Standard Model is
actually much more complicated than one might have expected on
esthetic grounds.  This is a subjective feeling, and long familiarity
with the Standard Model has perhaps dulled it for most of us, but this
point was keenly felt by physicists of the 1930's, whose intuitions
may be as valid as ours.

Still, it might be that our vacuum is at a ``symmetric'' point.  
It seems to us that this would have to mean ``symmetric'' in terms of
how it is situated in the structure containing all the vacua, so
one again needs some overall picture to make any such judgement.

The best argument we know against the idea of ``Vacuum Selection
Principle'' is simply the following.  Suppose we found a
well-motivated principle $x$, other than consistency, which predicted
string/M theory vacuum $X$.  Suppose we then determined by observation
that we actually lived in vacuum $Y$, different from $X$.
Would we conclude that string/M theory is wrong?  No, we would
conclude that principle $x$ was wrong.

Having provided more than enough philosophy for one physics paper,
let us turn to physics.

\newsec{Determining the set of string theory vacua}

A huge amount of work has been done on constructing string theory
compactifications, and at this point it is impossible to give a real
survey.  This would seem to call into question any claim that
one can discuss ``all'' string vacua in any concrete way at all.

Our claims to this effect will rest on two general hypotheses which we
feel are supported by existing work.  First, because of duality, our
constructions of string theory vacua are highly redundant: the known
vacua can be realized by many general types of construction, and in
many ways.  This leads to the idea that if one could extrapolate any
single class of construction out of the regime in which it is weakly
coupled, one would in fact reach all vacua.  Thus, if one can estimate
numbers and statistics of vacua taking into account the quantum
corrections, even in a qualitative way, a single class of
constructions could describe a finite $O(1)$ fraction of the vacua and
give a representative picture.

Second, by ``all'' vacua, we will mean vacua with $\CN=1$
supersymmetry, both Minkowski and AdS.  This is for the usual
theoretical and phenomenological reasons, but we will suggest a
further reason in section 4: by developing the ideas proposed there,
information about the supersymmetric vacua could provide information
about vacua which spontaneously break supersymmetry.

These hypotheses along with the relatively strong mathematical
technology one can apply to this case motivate basing our
considerations on constructions of $\CN=1$ compactifications using
branes in type
\I\ or type \II\ orientifolds of Calabi-Yau threefolds.  This includes
F theory, and known dualities relate this case fairly directly to the
heterotic on CY and (less directly) to the
M theory on $G_2$ manifold constructions, to the extent
that one can make a unified discussion at least of the problem without
quantum corrections.  The constructions differ greatly in how quantum
corrections arise, but we will discuss these effects in a different
way.

There are certainly constructions which have not been precisely
related by duality to this class, such as asymmetric orbifolds.  On
the other hand, there are ideas for how to do this; for example
asymmetric orbifolds have some relation to discrete torsion, and
discrete torsion brings in only $O(1)$ new choices.  At present it
seems reasonable to think that the set of constructions we understand
moderately well, or (if we are able to extrapolate to strong coupling)
even the subset of type \II\ brane constructions, describe an $O(1)$
fraction of the possibilities.

\subsec{What is a vacuum?}

Ultimately, our goal is to characterize or count nonsupersymmetric
vacua which might be candidates to describe observed physics.
We are free to make various more general definitions of ``vacuum''
along the way, as long as we have some idea how these are related
to our goal.

For us, a ``vacuum'' is a critical point $V'(\phi)=0$ of the effective
potential in a Lorentz symmetric four dimensional effective field
theory which might express low energy predictions of string/M theory
in some situation.  The assumption that the problem can be discussed
in effective field theory terms is motivated by the fact that so far
all observed physics can be so described.

Nonsupersymmetric vacua of a theory with any reasonable effective
potential will be isolated, unless the theory has a continuous global
symmetry.  There are arguments that string/M theory cannot have such
symmetries \banksglo, and granting this point, there appears to be no
ambiguity in counting the physically relevant vacua.

The problem of counting supersymmetric vacua is somewhat more open to
definition, as supersymmetric vacua can come in moduli spaces,
nontrivial fixed point theories might be counted with multiplicity,
and so on.  To some extent there are natural ways to make these
choices, based on the principle that we want a definition which
depends as little as possible on the details of the effective theory, 
as we will discuss shortly.

The general picture the reader should keep in mind is that a moduli
space of vacua in the early stages of analysis (before taking quantum
corrections into account) will be assigned a number which estimates
the number of vacua which will appear with all corrections taken into
account.  For example, one can argue very generally that the number of
supersymmetric vacua in a (globally) supersymmetric sigma model with
superpotential should be the Euler character of the target space, no
matter what the superpotential is, because this is the value of the
Witten index \wittenindex.  While one can easily find caveats and
exceptions to this statement, it might still be that the Witten index
is a good enough estimate of the number of vacua for our purposes.

The usual way that supersymmetric vacua are used to try to infer the
properties of nonsupersymmetric vacua is to postulate a supersymmetric
extension of the standard model and a hidden supersymmetry breaking
sector, and treat the effects of the latter on the former as an
explicit supersymmetry breaking.  We will not get into this level of
detail, but obviously this approach can be phrased in terms of
``tests,'' and one can ask what fraction of all models contain such a
hidden sector and what fraction contain a supersymmetric Standard
Model which is coupled to it in the right way.  What we will do
instead, is to discuss an approach which could lead to ``universal''
predictions for the ratio of nonsupersymmetric to supersymmetric
vacua, in section 4.

Having said this, we focus on the problem of counting $\CN=1$
four dimensional vacua with the Standard Model gauge group.
We recall the standard $\CN=1$ supergravity expression for the
potential (in units $M_{pl}=1$) \wessbagger;
\eqn\sgpotential{
V = e^K \left( \omega^{i\bar j} D_i W D_{\bar j} W^* - 3|W|^2 \right) + D^2
} 
with $K$ the K\"ahler potential, $\omega_{i\bar j}=\p_i{\bar\p_{\bar
j}}K$ the K\"ahler form, $W$ a holomorphic section of a line bundle
$\CL$ with $c_1(\CL)=-\omega$, and $D_i W = \p_i W + (\p_i K) W$ the
covariant derivative on sections of $\CL$.  $D^2$ represents the 
``D-flatness'' part of the potential.

A supersymmetric vacuum satisfies $D_i W=0$ and has cosmological
constant $\Lambda=-3e^K|W|^2$.  These vacua can be Minkowski or AdS in
the four dimensional space-time.  Both types are equally relevant for
our purposes, and we will not try to distinguish them in our counting,
for several reasons.

First, from the phenomenological point of view of dynamical supersymmetry
breaking, the cosmological constant will get additional corrections
after supersymmetry breaking, and the only reasonable condition to
enforce before taking these into account is $\Lambda > -c M_{susy}^4$.

Second, from the theoretical point of view of inferring the
distribution of nonsupersymmetric vacua from that for supersymmetric
vacua, clearly we need information about all supersymmetric vacua to
have any hope of doing this.

Finally, from the mathematical point of view, Minkowski vacua are much
harder to count.  While one has the advantage that the K\"ahler
potential drops out of the supersymmetry conditions, which are then
holomorphic, it turns out that this is far outweighed by the
disadvantage that the conditions $\p_i W = W = 0$ are more equations
than unknowns.  The existence of solutions to such over-determined
systems of equations is non-generic and depends on very specific
features of $W$; without exact results for $W$ there is no way to
decide whether such solutions exist, let alone how many might exist.

Since finding supersymmetric Minkowski vacua is not the physical
problem, and it is so difficult, we will not discuss it further.
Henceforth, unless otherwise specified, a supersymmetric vacuum is
a solution of $D_i W=0$, with no constraint on $W$ or $\Lambda$,
and of the D-flatness conditions.

\subsec{Estimating numbers of vacua after quantum corrections}

Although a fair amount is known about $\CN=1$ string/M theory
compactification in the weak coupling limit, and about supersymmetric
field theory at arbitrary coupling, we do not yet have a good
understanding of $\CN=1$ string/M theory at arbitrary coupling.

A lot of progress is being made on exact results, and as we discussed
in \stringscam\ it seems likely to us that within a few years we will
have usable exact results in string/M theory of the same character we
now have for $\CN=1$ supersymmetric gauge theory, namely a precise
description of the gauge symmetry, chiral field content and
superpotential for a large set of $\CN=1$ compactifications.  This
would allow us to put the discussion of supersymmetric vacua on a very
firm footing.

However, this may be overkill for the type of question we are asking
here.  We will use two simple ideas to get estimates of the number of
vacua after quantum corrections.

The first idea is simply to estimate numbers of vacua at weak
coupling, and use the fact that as we vary parameters, supersymmetric
vacua tend to move around but are not created or destroyed.  Most of
our weak coupling estimates will be combinatoric, counting
Calabi-Yau's, vector bundles, brane configurations etc.

Let us give as a general example of this, the problem of finding all
vacua of a supersymmetric quiver gauge theory, as discussed in
\trieste.  We consider quiver theories arising as world-volume
theories of D-branes in type \II\ strings; these have gauge group
$\prod U(N_i)$ all matter in bifundamentals $(\bar N_i,N_j)$, a
superpotential which is a sum of gauge invariant single trace
operators, and Fayet-Iliopoulos (FI) terms.  First, it can be shown that
all classical supersymmetric vacua can be constructed in terms of
vacua with unbroken gauge symmetry $U(1)$, by taking direct sums
of the gauge groups and matter configurations.  The vacua with 
unbroken $U(1)$ are called ``simple objects'' and in brane language
correspond to bound states of branes.  A configuration with $n$ copies
of a simple object has $U(n)$ unbroken symmetry, and so on.  

Let $n_{\vec N}=n_{N_1,N_2,\ldots}$ be the number of simple objects in
the $U(N_1)\times ...$ theory (for conciseness let us denote this
semisimple group as ``$U(\vec N)$''.
As a simple example, consider the
``$\CN=1^*$'' theory \ps, a $U(N)$ theory with three adjoint chiral
superfields and a superpotential $W=\tr X[Y,Z]+X^2+Y^2+Z^2$.  This has
the spectrum of bound states $n_k=1$, one for each representation of
$SU(2)$.

It is easy to write a generating function
for the number of all classical vacua:
\eqn\genclass{
\sum_{\vec N} N_{vac}(U(\vec N)) q^{\vec N} =
 \prod_{\vec N} \left({1\over 1-q^{\vec N}}\right)^{n_{\vec N}} .
}

Suppose furthermore that these simple objects  have no
remaining massless matter (are ``rigid''), 
then using the fact that
pure $SU(k)$ SYM has $k$ supersymmetric vacua it is easy to obtain
the generating function counting all vacua of the quantum theory.  It is
\eqn\genquant{
\sum_{\vec N} N_{vac}(U(\vec N)) q^{\vec N} =
 \prod_{\vec N}
   \left(1 + {q^{\vec N}\over(1-q^{\vec N})^2}\right)^{n_{\vec N}} .
}
For $\CN=1^*$, this gives the counting
$N_{vac} \sim c^{\sqrt{N}}$.  As we argue below, 
generic brane theories have many more bound states, and
the generic estimate of this type is $N_{vac} \sim c^N$.

Although this was a weak coupling argument, vacua in globally
supersymmetric theory will not be created or destroyed under
variations of the gauge couplings or variations of the superpotential
which do not change its asymptotics.  

The existence of supersymmetric vacua can depend on the
Fayet-Iliopoulos terms.  This behavior in the classical theory
is given by ``$\theta$-stability'' \king, according to which
certain solutions of $W'=0$ can be unstable for all values of the
Fayet-Iliopoulos terms, and typical solutions are stable within some
cone in the space of these parameters.  One can then use \genclass\
computed for some value of the FI terms, to get \genquant\ at that
value of the FI terms, and extrapolate the result to arbitrary 
coupling.\foot{There are exceptions to this rule which
can for example lead to supersymmetry breaking in the quantum theory.
Although this deserves more systematic study,
we suspect this is nongeneric (fortunately, there are other ways to break
supersymmetry).}

These ideas are simple but such results are not in themselves the answer,
because the gauge and superpotential couplings are not parameters in
string/M theory models.  All of these couplings are fields, and their
possible expectation values must be found by solving their equations
of motion, $DW/D\phi=0$.

This is a problem, but not a disaster, because if we start at a
critical point for all the other fields and follow the gradient of the
superpotential, a generic non-vacuum configuration must flow either to
a vacuum or to a boundary of moduli space.  If one knows that one of a
large class of such configurations flow to a vacuum, it is likely that
most or all do.  While heuristic, this idea justifies the claim that,
if one sector of the theory stabilizes couplings in another sector in
some generic cases, it will do so in an $O(1)$ fraction of cases.

These ideas can be made more precise by basing them on the Witten index
$\Tr\ (-1)^F$ in supersymmetric field theory \wittenindex.  Although
this is not literally the number of supersymmetric vacua, for theories
with isolated vacua
it is a lower bound, and probably a fairly accurate
one in theories with generic (complicated) superpotentials
and no unbroken $U(1)$'s.
There is a generalization of this index in effective supergravity theories
which counts vacua with signs, as we discuss elsewhere
\refs{\stringscam,\dougwit}.

Another generalization of the Witten index, to deal with unbroken
$U(1)$'s, can be motivated by returning to the
quiver gauge theories.  Consider theories with $U(N)$ gauge group;
all of these theories have $\Tr (-1)^F=n_N$, because
all of the vacua made up of more than one simple object
have unbroken $U(1)$'s and thus cancel out
of the Witten index \wittenindex.  This is because of the pairing of
the ground state with states obtained by applying the $U(1)$ gaugino 
operators $W_\alpha^n$.

The natural generalization of the Witten index to count these vacua
as well is to count
the operators in the chiral ring \cdsw.  This number
is very similar to $N_{vac}$
but counts each unbroken $U(1)$ at low energy with multiplicity $4$,
$$
N_{chiral\ ring} = N_{vac} \times 4^{N_{unbroken\ U(1)'s}} .
$$
This replaces \genquant\ with 
\eqn\genchiral{
\sum_{\vec N} N_{chiral\ ring}(U(\vec N)) q^{\vec N} =
 \prod_{\vec N}
   \left({1 + q^{\vec N}\over 1-q^{\vec N}}\right)^{2n_{\vec N}} .
}
The simpler form of this formula suggests that the number $N_{chiral\
ring}$ would have better formal properties than $N_{vac}$.  However
our considerations below will be too crude to benefit much from this
improvement.

Having seen the relevance of the Witten
index and these simple generalizations of it,
the second idea is to get ``topological'' formulas for them.
Let us give an example.

For theories with compact moduli spaces of vacua, the Witten index is an
estimate in the sense we want, {\it i.e.} a number which counts the
likely number of vacua after further quantum corrections.  This
follows if we assume that quantum corrections produce a superpotential
which is a generic function of the gauge invariant fields, because the
Witten index will be invariant under such a deformation.

For a (non-gauge) globally supersymmetric sigma model with target
space $E$, one can compute the Witten index by compactifying the
Minkowski space dimensions and reducing the discussion to
supersymmetric quantum mechanics.  Thus, we count a moduli space $E$
with multiplicity $\chi(E)$, its Euler character.

For (non-gauge) supersymmetric theory with a superpotential, vacua
are solutions to the equations $W'=0$.  Counting solutions to systems of
complex algebraic equations is a well understood problem and in a
certain sense the number is topological.  For example, for a generic
system of $n$ independent degree $d$ polynomials in $n$ unknowns,
there are $n^d$ simultaneous solutions.  If we assumed that the $n$
equations $W'=0$ were independent (of course they are not in general),
we would get an estimate, in terms of the degree of the polynomial $W$.

Both of these considerations can be incorporated 
in the following formula:
\eqn\sgindex{
N_{susy\ vac} = \int_\CC c_n(\Omega\CC \otimes \CL) ,
}
where $\Omega\CC$ is the complex cotangent bundle to the 
configuration space $\CC$, and
$\CL$ is the line bundle in which the superpotential takes its values.
For example, the sigma model case is $\CL$ trivial,
and this integral is the Euler number, while the degree $d$
superpotential can be treated by compactifying $\BC^n$ to $\BP^n$
and taking $\CL=\CO(d)$.  This does not give $d^n$ but the
supergravity index, which is comparable.

Heuristically, this formula is saying that to contain many vacua, a region
of the configuration space must have complicated topology, large
K\"ahler volume (in Planck units), or both.

This formula is well known to mathematicians as the general formula
for the number of critical points of $W$, given that $\CC$ is compact
and $W$ is holomorphic.  Although as written it assumes that $\CC$ is
a manifold, even this can be generalized \fulton.
In any case, it can easily be made precise if the configuration space
 $\CC$ is compact and $W$ is non-singular.

Unfortunately, these conditions are almost never satisfied by explicit
string/M theory superpotentials (and are literally impossible in
supergravity).  For noncompact $\CC$, one can still try to use
\sgindex\ by interpreting the integrand not as a topological class but
rather as an explicit form constructed from the K\"ahler metric
and curvature, and
we will suggest some justification for this idea in section 4.
This includes the case of $W$ non-single-valued, as one can go
to a covering space which makes $W$ single-valued, usually at the cost
of making $\CC$ noncompact.  

Now, granting \sgindex,
suppose our theory is composed of two sectors $F$ and $G$, where
$F$ has important supergravity corrections, while couplings in $G$ have strong
dependence on fields in $F$.  One can then approximate \sgindex\ as
\eqn\sgindexprod{\eqalign{
N_{susy\ vac} &=
  \int_{\CC_F} c_n(\Omega\CC_F \otimes \CL_F) \times
 \int_{\CC_G} c_n(\CL_G) \cr
 &\sim N_{susy\ vac\ F}\times N_{susy\ vac\ G} 
}}
since the number of vacua in $G$ does not depend on the parameters.

Thus, the formula \sgindex\ also supports the idea that to the extent
we can think of the theory as composed of two sectors, the number of
vacua will be roughly the product of the numbers in each sector, even if
couplings in one depend on fields in the other.  We suspect that many
exceptions to such claims can be constructed, but at the present state
of the art it is hard to make progress without making some such
claims, and we feel these arguments give some justification for them.

\subsec{Type \I\ and type \II\ orientifold models}

The class of models we will consider has been discussed in many works.
The prototype is to compactify ten dimensional type \I\ string theory
along the same lines as the original construction of quasi-realistic
$\CN=1$ vacua due to Candelas, Horowitz, Strominger and Witten \chsw.
This started from the heterotic string, but worked in the large
volume, weak coupling limit which is well described by $d=10$, $\CN=1$
supergravity/Yang-Mills theory with the standard anomaly cancellation
structure \greenschwarz, and thus their general discussion applies:
one compactifies to four dimensions on a Calabi-Yau threefold $M$,
choosing a gauge connection for a bundle $V$ on $M$ with structure
group $G\subset SO(32)$ and satisfying the anomaly cancellation
condition $c_2(V)=c_2(TM)$, and the Hermitian Yang-Mills equations.
One obtains an $\CN=1$ supersymmetric low energy theory with
gauge group $H$ the commutant of $G$ in $SO(32)$, and a spectrum of
charged chiral multiplets coming from massless adjoint fermion zero
modes on $M$.

One of the prime advantages of this construction is its relation to
algebraic geometry.  This is a very long story involving formidable
mathematics; at this point the general classification of allowed $M$
and $V$ is not known, but there are moderately effective techniques
for constructing examples and some understanding of the overall
picture, which we will try to use here.

We would like to base our discussion on the following claim:
there are three generalizations of this construction,
which if done in full generality, and extrapolated to strong
coupling, could lead to an $O(1)$ fraction of possible vacua.
First, we use the equivalence between
gauge field configurations and Dirichlet brane configurations,
exemplified by the relation between small instantons and D$5$-branes
\witteninst.  This has been greatly extended and generalized,
to the point where one can get a usable picture of the set of
all bundles $V$.  Second, one can apply ``generalized T-dualities'' to
obtain type \II\ orientifold constructions, in which perturbative
gauge symmetries can come from Dirichlet branes wrapping arbitrary
supersymmetric cycles.  Finally, one can turn on antisymmetric gauge
field strength fluxes.  The claim would be that general type \II\ backgrounds
with branes and fluxes cover an $O(1)$ fraction of the possibilities.

Although a fair amount is known about branes and fluxes separately,
unfortunately a complete description combining branes with fluxes is
not known at present, even in the supergravity (weak coupling and
large volume) limit.  This is an active subject of research and
the situation may improve before long.  At the present state of
knowledge, we are going to have to make some guesses as to how to do
this.

Let us go on and discuss the choices which enter this construction.

\subsec{The choice of Calabi-Yau}

Basic introductions to Calabi-Yau compactification can be found in 
\refs{\gsw,\greene}.

Construction of Calabi-Yau threefolds has been much studied
and there is a subset which in a sense has been classified, the
hypersurfaces in toric varieties.  In \kreusch, Kreutzer and 
Skarke classify an appropriate type of ``reflexive polyhedron''
which can be used to construct such a \CY3, and
show that the number of these is $N^\ub_{CY_3} = 473,800,776$.
Since distinct polyhedra can lead to the same \CY3, this number
is an upper bound for $N_{toric\ CY_3}$ (thus our notation).
On the other hand, \CY3's with
distinct Betti numbers $(b_{1,1},b_{2,1})$ 
are clearly distinct; the number of distinct
pairs which appear is $N^\lb_{CY_3}=30,108$ which is
a lower bound.  There are pairs of distinct
\CY3's with the same Betti numbers, so this bound is not sharp either.

Plotting the Betti numbers produces a diagram (the ``shield'') which
obviously has structure, supporting the idea that this is at least a
natural subclass of \CY3's.  Within this class, the Euler character
$\chi(M)=2(b_{1,1}-b_{2,1})$ satisfies the bound $|\chi(M)|\le 960$.
The number of distinct $b^{1,1}$'s
for a given $\chi$ is roughly $2\chi$ for $\chi<320$,
and decreases for larger $\chi$.  Some patterns in their other
topological invariants are observed in \schimmrigk.

Unfortunately it is not known whether all \CY3's are of this type.
Indeed, mathematicians still debate whether there are finitely many or
infinitely many distinct $M$.\foot{More precisely, we want the number
of components of the moduli space of birational equivalence classes of
complex \CY3's.}  Most seem to believe that the number is finite.  The
evidence for this, such as it is, is that (1) mathematicians know no
example which is not a toric hypersurface, and (2) one can start with
an $M$ with (say) $\chi=960$ and try to increase $\chi$ by an extremal
transition; so far this has not led to new examples \markgross.

Clearly for present purposes one can only assume that the list of
\kreusch\ is representative, and use the bounds we just obtained as
the estimated number of possibilities.  This type of uncertainty will
plague our discussion, and this estimate should be refined, but far
greater uncertainties await us.  From now on, when we say that ``we
will assume construction $X$ is representative,'' we mean that in the
discussion in section 5, we will assume that the choices involved at
that step lead to an $O(1)$ fraction of the possibilities.

Finally, these discrete choices do not uniquely characterize the Ricci
flat metric on $M$: one has additional continuous parameters or
``moduli.''  On a general level this is well known: there are
$b_{1,1}+b_{2,1}$ moduli, each leading to a singlet chiral superfield in
the low energy effective Lagrangian.

Quite a lot is known about the global structure of these moduli spaces
and even explicit metrics are known, at least in the weak coupling and
large volume limit.  This comes from considering the related type \II\
compactifications on CY with $\CN=2$, $d=4$ supersymmetry, and using
``special geometry.''  The simplest picture, with the broadest
applicability,
comes from considering complex structure moduli space, since there are
$\CN=2$ type \II\ compactifications for which this metric is exact.

We refer to \candelas\ for a detailed study of the ``mirror quintic''
\CY3 complex structure moduli space, with one chiral superfield, as
perhaps the simplest example with almost all the qualitative features
of the general case.  A simpler example with most of the features is
the complex structure moduli space of the torus $T^6$,as discussed in
many references \refs{\moore,\kst}, while the mathematical technology
for the general case is discussed in \cox.

A complex torus $T^{2n}$ can be defined as the space $\BC^n$ quotiented
by a $2n$-dimensional lattice $\BZ^{2n}$.  Explicitly, let $u^i$ be
coordinates on the torus; we identify $u^i \sim u^i + m^i + Z^{ij} n_j$
for all $m,n\in\BZ$, with
$Z^{ij}$ a complex matrix with positive definite imaginary part;
call the space of these $Mat^+_n(\BC)$.

While this construction determines the complex structure on $T^{2n}$, 
the relation is not one-to-one: two lattices $(1,Z)$
related by an $SL(2n,\BZ)$ transformation
lead to tori with equivalent complex structures, related by a large
diffeomorphism which acts nontrivially on the periods.  Thus the moduli
space of complex tori $\CM_c(T^{2n})$ is a quotient $Mat^+_n(\BC)/SL(2n,\BZ)$.

The physical metric on this moduli space, which appears in the
supergravity kinetic term, is the Weil-Peterson metric on this moduli
space of flat metrics (this is a fancy way to say, the metric which
arises from straightforward Kaluza-Klein reduction).  It has K\"ahler
potential
\eqn\tsixK{
K_Z = -\log \det \Im Z
}
and constant negative curvature.  Constant negative curvature is
special to this example, but negative curvature is a very general feature
of Weil-Peterson metrics.

One of the important qualitative features of this moduli space, and other
\CY3 moduli spaces, is that it has finite volume in the K\"ahler metric
\hornemoore.  This is despite the fact that boundaries can be at
infinite distance.

Finally, another important quantity which can be computed for very
general \CY3's as a function of the moduli
is the vector of periods of the holomorphic three-form,
\eqn\periods{
\Pi^i = \int_{\Sigma_i} \Omega .
}
Here $\Sigma_i$ is a basis for $H_3(M,\BZ)$.
For $T^6$, for example, these are $1$, $Z_{ij}$, 
$(\det Z)(Z^{-1})^{ij}$ and $(\det Z)$.
These enter in flux superpotentials and in black hole entropy
calculations, to name two applications we will call on.  Note that
they transform non-trivially under $SL(6,\BZ)$ and are thus
not single-valued on $\CM_c(T^{2n})$, but only on 
$Mat^+_n(\BC)$.

\subsec{Bundle and brane configurations}

This is another long story from which we will try to extract a general
picture by combining various dualities and brane arguments with
considerations in algebraic geometry.  We will base this on four
general approaches: the ``large volume'' approach involving study of
holomorphic bundles, the bound state/derived category approach,
enumerative results on curves, and the spectral cover/T-duality
approach.  Our primary question is still whether the number of
possibilities is finite and whether we can estimate it.  We will also
need to decide what fraction of constructions are likely to produce
Standard Model gauge groups and chiral matter content.

In the large volume approach, the problem reduces to that of finding
solutions of the hermitian Yang-Mills equations on $M$.  Among the
many general references on this problem are \donkron, \trieste, and
\dougzhou\ which will summarize some general facts relevant for
superstring compactification.  In particular, it is known that there
is a $n$-dimensional region within the $n$-dimensional lattice of
Chern classes (or brane RR charges, or K theory) for which stable
bundles exist.  By tensoring with a line bundle, one can always set
$c_1(V)=0$; the region is then roughly characterized by the bounds
$c_2(V) > 0$ (more precisely, one has the Bogomolov bound) and bounds
on the remaining invariant $c_3(V)$, which can only take finitely many
possible values.
Unfortunately no effective bound is known,
but various general considerations point to a bound of the order
$|c_3(V)| < C\cdot\chi(M)$.  Furthermore, the resulting moduli spaces
of bundles (of fixed topological type) are algebraic, meaning
essentially that they can be defined by a finite system of equations
in some (large dimensional) projective space.  Although they need not
be manifolds, this is good enough to believe that there will be
finitely many vacua after quantum corrections.

Although not terribly concrete, these are at least good finiteness
results for the large volume, weak coupling limit.  Once we leave
this limit, the situation is less clear.\foot{
For example, let us consider
the dual heterotic M theory picture.  Here we can add five-branes,
wrapping effective cycles $\Sigma$ satisfying
$c_2(V) + [\Sigma] = c_2(TM)$.
There is still an argument for finiteness of the number of solutions
to this more general problem.  It is that the ``effective'' condition
in the choice of five-brane, which along with the Bogomolov inequality
tend to make both bundle and brane contributions to this formula
positive (physically this is to say that the branes and instantons
must be BPS).  the number of choices is finite here as well.
However, the Bogomolov inequality only bounds a single component of $c_2(V)$,
namely $\int_M c_2(V)\wedge \omega$, leaving open the possibility that
there are infinite sequences of stable bundles with the other
components of $c_2(V)$ running off to negative infinity.  No such
example is known, but it could be that they exist and are ruled out on
other physical grounds (we will discuss analogous examples later).
I thank Richard Thomas for a discussion on this point.}

Making a comparable discussion
for type \II\ and orientifolds requires using the general relations
between vector bundles on Dirichlet branes and combinations of
D-branes discussed on a basic level in
\refs{\polchinski,\johnson}.  This allows turning the problem of
classifying bundles into that of understanding moduli spaces of brane
configurations.

A general result which can be derived from the index theorem at large
volume, but applies to all approaches, is the following.  Consider a
configuration of $N_i$ branes of type $B_i$ and $N_j$ of type $B_j$
in type \II\ string theory,
where each brane is ``simple,'' i.e. comes with $U(1)$ gauge symmetry.
These could be branes wrapping different cycles, or carrying different
gauge connections, or whatever.  In any case, the net number
of chiral multiplets arising from open strings between these branes
is given by a bilinear form in their RR charges (or K theory classes),
the ``intersection form,'' which we denote
$$
N_{(\bar N_i,N_j)} - N_{(\bar N_j,N_i)} =
I_{ij} = -I_{ji} = \vev{B_i,B_j} .
$$
Simple explicit formulas can be found for this form in all approaches.

There are similar formulas for type \I\ and orientifolds
\refs{\sagnotti,\csu}, involving the orientifold action and the class
of the fixed plane.  Rather than use these formulas, we are going to
use a simpler description of orientifolding \dm: given a $U(N)$ quiver
theory in which the intersection numbers and ranks of gauge groups
have symmetry under a $\BZ_2$ action on the nodes, one can restrict
attention to gauge fields and matter configurations which are
invariant under the symmetry (with suitably chosen signs).  
Some of the geometric definitions of orientifolding using ``image
branes'' can be shown to reduce to this,
and we will assume that this construction is representative,
within the context we discuss below.

This allows us to find the massless matter spectrum for simple
combinations of branes.  However, this only scratches the surface of
the problem as general holomorphic bundles correspond to general bound
states of branes, typically with very complicated moduli spaces, as
one would expect for classical moduli spaces of supersymmetric vacua
of gauge theories with generic superpotentials.  
While there has been much mathematical
work on the problem, it is not easy to
explicitly describe the bundles and moduli spaces even on the simplest
threefold, projective space $\BP^3$.  The Calabi-Yau case is 
similar but harder.

On penetrating the language and other barriers,
one finds that much of this mathematics
turns out to be based on ideas which have
relatively simple physical translations, which we refer to as the
``bound state/derived category approach,'' as discussed in \trieste.
Recent work has led to a fairly good understanding of this translation
in type \II\ theory, which in broad terms can be summarized in the
claim that a supersymmetric brane configuration in type \IIb\ at weak
coupling but arbitrary K\"ahler moduli is a $\Pi$-stable object $E$ in
$D(\Coh M)$, the derived category of coherent sheaves.  The next step
in a systematic approach to the models under discussion is to classify
possible orientifoldings $\Omega$; these are in a sense $\BZ_2$
automorphisms of $D(\Coh M)$ which it is plausible to believe are
obtained by conjugating the type \I\ $\Omega$ by the action of
Fourier-Mukai transforms (T-duality).

What this means in more physical terms is the following.  We start
with a small set or ``basis'' of elementary branes, at least one for
each K theory class on $M$, and try to describe all branes as bound
states of these elementary branes and their antibranes.  This is done
by deriving the joint world-volume theory of the collection of branes;
each bound state is then a supersymmetric vacuum of this theory.

The power of this approach comes from the fact that one can find bases
with simple world-volume theories.  One looks for ``rigid'' branes,
meaning those without world-volume adjoint matter, chosen so that any
pair within the set has only open strings of a single charge (in
quiver language, a single orientation of arrows), as the
superpotential for up to two branes is then forced to be zero by gauge
invariance, and can be computed systematically for more branes.  The
set of ``fractional branes'' in an orbifold or Gepner model \ddg\
provides an example; one can find other bases by applying Seiberg
duality to this one, and the general picture is that any bound state
is a bound state of branes (not antibranes) in one of these preferred
bases.

These Seiberg dualities can have various physical interpretations
depending on the gauge couplings; at weak coupling one is simply using
different bases of branes to describe the same bound states, while
more generally couplings can flow and different dual pictures can be
valid at different energy scales \refs{\cachdual,\fiol}.  One would
need to take this into account to decide which of these Seiberg dual
theories are physically dual, and which are different at the scale of
supersymmetry breaking (below which the duality is inoperative).

The simplest examples are the hypersurfaces in weighted projective space,
a subset of the toric hypersurfaces with $7,555$ elements.
Some of these can be defined in string theory as Gepner models.\gepner\ %
In these models, there is a preferred basis of ``fractional branes,'' and a
simple description of their intersection form.  A Gepner model is
essentially a $\BC^5/\BZ_K$ Landau-Ginzburg orbifold model,
characterized by a choice of $\BZ_K$ action on $\BC_5$ and some
continuous parameters (superpotential and FI terms).  The $\BZ_K$
action can be characterized by a choice of five integers $a_i$ which
sum to $K$ and satisfy the constraint $K/a_i\in\BZ$.
One can compute the Betti numbers from this data, 
and one gets roughly $b^{1,1} \sim K$.

As discussed in \dd, this model has $K$ fractional branes $B_i$
with $0\le i < K$, whose intersection form is simply expressed as
\eqn\intnum{
\sum_j q^j \vev{B_i,B_{i+j(\mod K)}} = \prod_{n=1}^5 (1 - q^{a_n}) .
}
Enough is known about the superpotential and other data of this theory
to get moduli spaces of many simple bound states, as discussed in \dgjt.
It is known how to get similar results for general toric
hypersurfaces, although this remains to be done explicitly \katzcom.

To use such a theory in string theory, one must choose an
orientifolding and enforce anomaly cancellation.  The orientifoldings
which are simple in the quiver language are the ones we described
above which project on configurations which are invariant under a
$\BZ_2$ reflection of the quiver; there is a partially understood
relation between this and the geometric definitions of orientifold.
In any case, the resulting anomaly cancellation condition is the same
as that in the large volume type \I\ and
\IIb\ geometric orientifold constructions \roemel.  For compact
CY, the fractional branes provide an overcomplete basis for the K theory
so this determines the numbers of fractional branes up to a few
adjustible parameters.  These conditions can be worked out explicitly,
but we will only call upon their qualitative form: in examples, 
the solutions have many but not all of these numbers (so, ranks
of gauge group) non-zero, of order $O(10-100)$.  

This of course does not mean that the ranks of gauge groups need be
$O(10-100)$ as there is a lot of charged matter available to break the
gauge symmetry.  The basic assumption we will make in using these
theories in section 5 is that one can usually ({\it i.e.}, in an
$O(1)$ fraction of models) use this freedom to break this to a
specified subgroup while preserving supersymmetry; in brane language
forming bound states between some subset of the fractional branes.  If
so, then an $O(1)$ fraction of models which potentially contain the
Standard Model (by focusing on a subset of the anomaly cancelling
branes), will contain it.

This is nontrivial and often considered the hard part of the problem;
it involves details of the superpotential and D-flatness conditions
and is not always true.  Furthermore, even when it is true, the bound
states often involve fields with string scale vevs (since this sets
the scale of the FI terms), and one might worry about whether field
theory is justified.

Our main reason for nevertheless making such a claim is that the
relation between branes and geometric objects (bundles, objects in
$D(\Coh M)$ etc.) relates this question to questions such as whether
there is a stable bundle of the required topological type (and similar
questions) for which there is independent information.  As we
discussed earlier, this is true for some finite region in charge
space, and the anomaly cancellation conditions are usually such that
$c_2(V)>0$, so it seems reasonable to expect an $O(1)$ fraction of
solutions to the anomaly cancellation conditions to sit in this
region.  We should say that far more testing of this claim is possible
and would be desirable.

Another relation which lends some support to this idea is the relation
between brane configurations (at weak coupling) and black holes.  
This is obtained by reinterpreting the space-filling branes as
particles in the related \IIa\ string theory (formally, T-dualizing
the Minkowski space dimensions).  From the world-volume
point of view, this is reducing the supersymmetric gauge theory to
quantum mechanics, but many qualitative aspects such as stability and
supersymmetry breaking, and the estimate $N_{vac} \sim \chi$ (the
Euler character of the moduli space), are preserved under this.

By going to the strong string coupling limit, such a brane system
turns into a rather different system, a black hole.  The by-now
familiar idea that the entropy of a D-brane world-volume theories
should match that of a black hole in supergravity \sv\ provides a very
different way to get such estimates, by using the attractor mechanism
in supergravity.  It also provides a very different way to show
that certain brane configurations can form stable bound states:
if this entropy is non-zero, such a configuration must exist
\refs{\moore,\denef}.  This confirms the idea
that there is a finite volume region in the charge space for which
such bound states exist.\foot{
One should note that the black hole entropies,
which tend to go as $c^{N^2}$, are not in general a good estimate
for $N_{vac}=\chi$, as the black hole states are expected to
contribute to $\chi$ with signs.  In fact one typically obtains
$\chi\sim c^N$. \vafabh}

We have now laid out a certain style of analysis of the quiver gauge
theory of fractional branes at the Gepner point, on which we will base
the discussion in section 5.  These are however only
a subset of the rigid branes.  More rigid
branes can be obtained by performing Seiberg dualities on the
original quiver theory.  In the well understood examples (orbifolds),
all the rigid branes can be obtained this way, and we will assume
these are representative.  Seiberg duality acts
in a relatively simple way on quiver theories with no adjoint matter
\refs{\sei,\bea,\cachdual,\fengdual,\bd};
one picks a node of the quiver and applies the
duality of \sei\ to this node, treating the other gauge groups
as non-dynamical.  One can check that dualizing the $n$'th node
acts on the intersection numbers as
\eqn\seiint{
I_{ij} \rightarrow I_{ij} - 2 I_{in} - 2 I_{nj} + I_{in} |I_{nj}| .
}
This provides a large number of quiver theories from a single CY.
It is not known how many are distinct; the naive estimate $2^K$
which comes from allowing duality on each node independently is
clearly an overestimate in the known examples.  There is a (still
not well understood) relationship between these duality actions,
the CY monodromy group (acting on K\"ahler moduli space) and the
``phase structure'' of \agm, which suggests that the number should
be comparable to the number of ``phases'' of the model, which
is probably a low power of $K$.  It would be nice to have similar
results for the orientifolded theories, but Seiberg duality for these
has not been studied systematically.

We will use these results below to estimate numbers of brane constructions
which can realize the Standard Model.  Let us conclude by
briefly discussing the final two classes of construction.  A particularly
simple class of models is one in which the orientifolding fixes only
curves, so that anomaly cancellation can be accomplished using only
D$5$-branes wrapped on curves.  An example is given in \ach.
An advantage of this type of model is
that there is a highly developed technology for counting configurations
of curves on Calabi-Yau's, the ``original'' mirror symmetry
technology.  By wrapping branes on curves,
this can be used to count numbers of vacua directly.  The result
can also be interpreted as an Euler 
character of the moduli space of curves of the given degree.
This leads to an estimate of the form $c^N$, exponential in the
charge of the branes (degree of the curves).

Finally, we should mention the spectral cover construction
\refs{\donagi,\fmw}, which is a
very powerful and general construction of bundles on elliptically
fibered \CY3's.  The physical idea here is simple: by T-dualizing on
the fiber, a configuration of D$9$-branes carrying a very general
bundle can be turned into a configuration of D$7$ and lower
dimensional branes wrapping the base.  Generically, these D$7$-branes
will sit at different places in the fibration, in which case the
bundle data on these is simply the choice of a line bundle on each
D$7$-brane.  The only {\it a priori} condition on the bundles one gets
out is that the T-dual of the class of the D$9$ on the dual fibration
must be absent; even this restriction can be overcome by further
generalization (physically, taking bound states of the result with 
D$7$-branes in this remaining class).
Counting vacua in this type of construction thus boils down to counting
the configurations of D$7$-branes of a particular charge.

A simpler example which is related by T-duality to the ones we
discussed is to take space-filling D$3$-branes at points in $M$, as in
\kst.  If no superpotential is generated, the moduli space of $N$ such
branes is obviously $M^N/S_N$.  This space has Euler character roughly
$(\chi(M)+N)!/\chi(M)!N! \sim 4^N$ if $N \sim \chi(M)$, 
neglecting the singularities where branes
coincide.  Treating these singularities as quantum theories with
enhanced gauge symmetry can lead to larger estimates, but still
$O(c^N)$.

\medskip

This was a long subsection, so let us recap some of the main points.
First, the evidence seems consistent with the idea that there are
finitely many choices at this stage, which is important, as one
expects a finite fraction of these configurations
to match the Standard Model.  Second, a generic gauge theory
with $N$ branes would be expected to contribute a $c^N$
multiplicity of vacua.  Finally, we have systematic techniques for
constructing large numbers of configurations, which we will use
later to discuss the difficulty of realizing the Standard Model.

\subsec{Flux contributions and the cosmological constant}

Besides metric, Yang-Mills and brane degrees of freedom, string/M
theory contains various $p$-form gauge fields.
All of the well understood compactifications can be generalized
by turning on background flux for these gauge fields, as first
discussed in \stromtor\ and generalized in many works
(a few are \refs{\psflux,\beckerG,\dasgupta}).
This flux leads to a potential
energy (the ``flux potential'') which can be explicitly computed
in many examples, at least at weak coupling.

This work has led to two important physical ideas, which we will
review and build upon.  First, since the potential energy from the
flux depends on the moduli of the internal manifold in a fairly
complicated way, one expects it to have isolated minima; in other
words the moduli are stabilized.
This idea has a long history; recent work has focused on the use
of exact results for the flux potential, and 
in work of Giddings, Kachru and Polchinski
\gkp, Acharya \achflux\ and Kachru, Kallosh, Linde and Trivedi \linde,
it has been shown that moduli can be stabilized at finite coupling and
volume, as we discuss shortly.

Second, Bousso and Polchinski \bousspol\ have suggested that the large
number of independent flux contributions can lead to a large set of
vacua with a closely spaced spectrum of cosmological constants, so
that it becomes likely that vacua exist with acceptably small
cosmological constant.  Related ideas were proposed in \feng.

By the rules we stated in the introduction, this can count as a
solution to the cosmological constant problem, because we are not
insisting that there be a mechanism or selection principle which picks
out the observed case.  One still needs to check that a vacuum with
the appropriate $\Lambda$ exists, is metastable, can have reasonable
cosmology, and so on.  Since we observe $\Lambda>0$, the constraint of
metastability seems to be mild, because most likely decays are to AdS
vacua, which by general considerations of quantum gravity are highly
suppressed or impossible \refs{\coleman,\banks}.

Thus, fluxes seem to provide concrete candidate solutions to some
important problems, as well as potentially dominating the other
possible types of vacuum multiplicity.

We start again with a broad outline to make some basic points.
Let $C^{(p)}$ be a $p$-form gauge field, and $F^{(p+1)} = dC^{(p)}$ be
its field strength.  Lorentz invariance of the vacuum is preserved
either by electric $4$-form flux in Minkowski space, possible for
$p\ge 3$ if there are $p-3$ cycles, or by taking a magnetic $p+1$-form
flux in the internal space.  These are interchanged under duality
$*F^{(p+1)}=F^{(D-p-1)}$, so by considering both dual representations
of the gauge field, we can restrict attention to magnetic fluxes.

In a minimal energy flux configuration, $H^{(p+1)}$ is a harmonic form,
and is characterized by the cohomology class
of the field strength
\eqn\fluxquantum{
N = {1\over e}[F^{(p+1)}] \in H^{(p+1)}(M,\BZ) ,
}
a $J=b_{p+1}$-component vector.
For physical reasons discussed in \refs{\polchinski,\bousspol},
the flux must satisfy a quantization condition
$\int_\Sigma H^{(p+1)} = c N/M^p$; {\it i.e.} 
$e = c/M^p V_{p+1}$ in \fluxquantum, where $V_{p+1}$ is the volume
of the wrapped cycle, and $M$ a fundamental scale, typically $O(M_P)$
where $M_P$ is the higher dimensional Planck scale.

The flux contributes its potential energy to the effective potential:
\eqn\fluxpotential{
V_{flux\ 1} = \int_M H \wedge *H \sim M_{pl}^2
 {N^2 c^2\over M^{2p} V_{p+1}^2} ,
}
where the four dimensional Planck scale is related to the $D$
dimensional Planck scale as $M_{pl}^2 = M_{P}^D V_M$.
This formula might be modified by gravitational backreaction effects,
stringy and quantum corrections.

Let us now review the discussion of Bousso and Polchinski.  Following the
ideas of Brown and Teitelboim \brown, to get a small
cosmological constant, one assumes that the effective potential
is the sum of a large negative term $-\Lambda_0$ and the flux contribution
\fluxpotential.  Although flux quantization forces \fluxpotential\ to
take one of a discrete set of values, if there are enough distinct choices
of flux whose energy spacings are small compared to $\Lambda_0$, it will
be likely to find a discrete choice with cosmological constant within
the experimental bound.  

The necessary condition for this can be stated most simply in terms of
the number distribution for vacua with a given flux potential $V$,
a measure $d\mu(V)$ defined by
\eqn\fluxmeasure{
d\mu(V) = \sum_{T\in theories} \delta(V-V(T)) .
}
We will discuss this type of ``ensemble observable'' in more depth in
section 4, but in this simple example the definition should be clear.
In terms of this distribution, the condition is then simply
\eqn\cccond{
1 << \int_{\Lambda_0+\Lambda_{min}}^{\Lambda_0+\Lambda_{max}} d\mu(V) 
}
where $(\Lambda_{min},\Lambda_{max})$ are the experimental bounds
on the cosmological constant.  Evaluating \fluxmeasure\ using
\fluxpotential\ gives approximately
$$\eqalign{
d\mu(V) &= \sum_N \delta(V-{M_{pl}^2 N^2\over M^{2p} V_{p+1}^2}) \cr
&\sim {\Vol(S^{J-1})\over 2}\int dN^2 N^{J-2}
 \delta(V-{M_{pl}^2 N^2 \over M^{2p} V_{p+1}^2}) \cr
&\sim {\Vol(S^{J-1})\over 2}
 \left({M^{2p} V_{p+1}^2 \over M_{pl}^2}\right)^{J/2} V^{J/2-1} dV.
}$$
Replacing the sum with an integral is reasonable when \cccond\ is true,
so one gets a condition
\eqn\naivebp{
1 << \Delta\Lambda
 \left({M^{2p-D} V_{p+1}^2 \over V_M}\right)^{J/2} \Lambda_0^{J/2-1} .
}

One can then combine this condition with constraints on the other
quantities entering \fluxpotential\ to get a picture of the class
of models in which this works.  First, if the geometry of the internal
space is ``not too anisotropic'' ({\it i.e.} we are away from limits
or singularities in moduli space), we can take
$V_{p+1} \sim V_M^{(p+1)/(D-4)}$.  \naivebp\ then
reduces to
$$
{\Lambda_0\over\Delta\Lambda} << (\Lambda_0 V_M^{4/D})^{J/2} .
$$

The appropriate bound on $\Lambda_0$ is not at all obvious.  Indeed,
it is not immediately apparent where negative contributions to the
vacuum energy will come from, and once we find them, we will face the
potential problem that we will find a series of vacua in which
$-\Lambda_0$ can become arbitrarily negative, leading to an infinite
set of vacua and complete loss of predictivity.  

The simplest guess is $\Lambda_0 \sim M_{pl}^4$.  This works well in
large extra dimension scenarios, as $V_M$ is large.  In the
traditional weakly coupled string models, with $V_M \sim
(\alpha')^{D/2}$, one finds that $\Lambda_0 V_M^{4/6}$ is small, but
since $J=b_{p+1} \sim 100$ is typical for \CY3's, at first sight this
seems viable.  On the other hand, as will be clear below, the
flux contributions which cancel $-\Lambda_0$ typically lead to
supersymmetry breaking at a scale $\Lambda_0^{1/4}$,\foot{I thank
Shamit Kachru for emphasizing this point.}, and $\Lambda_0 \sim M_{pl}^4$
is not acceptable from this point of view.

In any case, this discussion demonstrates the possibility of large
multiplicities of vacua with cosmological constant uniformly
distributed near zero, and thus a potential solution to this problem
by our rules.  The discussions in \refs{\brown,\bousspol,\feng} attempted to
go further and explain the observed low value as the result of a
natural decay process involving nucleation of domain walls which
source the flux and lower the vacuum energy, down to some minimum
positive value.  They found that this mechanism is difficult to
realize as tunneling rates between flux vacua, even in the best case,
tend to be too small.  We will not insist on this or on the simple 
form \fluxpotential, but only on \cccond, which could be realized by
many types of degeneracy, including those in which the relevant vacua
had wildly different microscopic origins.

\subsec{A finiteness conjecture}

The previous discussion was somewhat simplistic as it ignored the fact
that the moduli of the internal cycles which minimize the true
effective potential in fact depend on the fluxes.  This was
inessential to the main point of \bousspol, but as a next step needs to be
taken into account.

The most obvious question this dependence raises is that the form of
\fluxpotential\ admits the possibility of sequences of vacua in which
both fluxes $N$ and volumes $V_{p+1}$ run off to infinity in a
correlated way, such that the cosmological constant stays finite.  If
so, it would simply not be true that the number of vacua is finite.

In fact the existence of infinite lists of vacua is well-known in
models with more supersymmetry.  The most famous example is perhaps
$S^5$ compactification of the \IIb\ string, which has infinitely many
vacua, parameterized by the number $N$ of quantized units of five-form
flux.  In this case, the radius of $S^5$ is proportional to $N^{1/4}$,
so this family runs off to large volume and small cosmological
constant.  In \trivedi, a similar series of $T^6$ flux compactifications
was found.  These are non-supersymmetric no scale compactifications, but
there seems no reason not to expect similar supersymmetric examples.

According to our rules, this is not a problem if none of these
infinite series of vacua look like the real world, and if volumes
of cycles $V_{p+1}$ run off to infinity, one is certainly tempted to
say that the total volume $V_M$ will as well, and the four
dimensional Planck scale $M_{pl}$ will run off to infinity.

Thus, in the absence of further constraints, the predictivity of
string/M theory depends on the
conjecture that the number of consistent flux vacua with
cosmological constant $|\Lambda| < \Lambda_{max}$, a bound we choose,
and compactification volume $V_M < V^\ub$, a upper bound,
is finite.  In examples (we will discuss one shortly), one also has
constraints on the fluxes from anomaly cancellation, which depend on
other input (numbers of branes and topology of the \CY3); let us
denote this input as $[B]$.

Then, the conjecture is that the total number of vacua, summing over all
allowed values of the flux,
\eqn\fluxconj{
N_{flux\ vac}(\Lambda_{max},V^\ub,[B]) \in \BZ
}
is finite.  We would conjecture this for any type of vacua, but
supersymmetric vacua (AdS and Minkowski together) would be the simplest
case to check.

This conjecture is not proven in any case we know of and might need to
be further refined.  One possible refinement would be to replace the
bound on total volume $V_M$ with bounds more directly related to
observation, because the appropriate bound on $V_M$ is very different
in the traditional dynamical supersymmetry breaking scenarios, and in
the ``large extra dimension'' scenarios.  This would be worth
developing, but in either case physics does place an upper bound
on $V_M$.

{}From what we have said so far, the most obvious way this conjecture 
could fail would be to find a series of models in which $-\Lambda_0$,
the vacuum energy at ``zero flux,'' became arbitrarily negative,
because one expects to be able to add fluxes to compensate it.

\subsec{Exact flux potential in \IIb}

An exact result in the large volume limit of \IIb\ string
compactification can be obtained by writing the potential in terms of
the Gukov-Vafa-Witten superpotential \gvw\ and using the special
geometry results we cited above to compute the periods of the CY.
This superpotential is a function of the CY complex structure moduli
$z^i$ and the axion-dilaton $\tau$.  We will also need to discuss
K\"ahler moduli; let $\rho$ be a K\"ahler modulus.  

We then have
\eqn\gvwii{
W = \int \Omega \wedge (F_{RR}^{(3)} + \tau H_{NS}^{(3)}) ,
}
where $F_{RR}$ and $H_{NS}$ are the Ramond-Ramond and Neveu-Schwarz
three-form field strenghs of \IIb\ string theory.
Using the quantization \fluxquantum, this can also be written
\eqn\gvww{
W_{flux} = \sum_i \Pi_i (N_{RR}^i + \tau N_{NS}^i) ,
}
with $N^i\in\BZ$ and $\Pi_i$ the periods defined above.
At large volume, the K\"ahler potential can be
obtained by KK reduction; it is (in terms of the \CY3 moduli space
K\"ahler potential $K_Z$, as in \tsixK)
\eqn\Krho{
K_{\tau,\rho} = K_Z - \log \Im \tau - 3 \log \Im \rho .
}
Using $K$ and $W$ in the standard $\CN=1$ supergravity
expression \sgpotential, one obtains the flux effective potential 
$V_{flux\ 2}$.

One cannot choose arbitrary fluxes; there is a constraint from
anomaly cancellation.  This will only work for \IIb\ orientifolds, 
for the simplest case of $O3$ planes it requires a tadpole
cancellation condition of the form \gkp %
\eqn\anomalyflux{
{1\over (2\pi)^4\alpha'^2}
 \int H_{NS}^{(3)} \wedge F_{RR}^{(3)}
= \vev{N_{NS},N_{RR}} = K - N_{D3}
}
where $\vev{N_{RR},N_{NS}}$ is the intersection form, $K$ is a positive
integer (the orientifold tadpole) and $N_{D3}$ the number of 
space-filling D3 branes (which must be nonnegative for supersymmetry).
Furthermore, one can show \refs{\gvw,\kst}\ that $\CN=1$ supersymmetry
implies that this number is non-negative.  This combination of facts 
gives a bound valid for supersymmetric vacua,
$$
0 \le \vev{N_{NS},N_{RR}} \le K .
$$
One might think that this provides an {\it a priori} bound on the
total flux, which would be very helpful in proving that the
number of vacua is finite.  This bound may be necessary, but
as we discussed above, we believe one needs to place
additional conditions on the flux vacua to get a finite number.
In particular, the form which appears in this bound is an
indefinite form (as is any bilinear form in two independent vectors),
so an infinite number of choices of flux satisfy this bound.  

The formula \gvww\ in itself is rather abstract; one needs to know
something about the behavior of \CY3 periods $\Pi^i$ to have any
intuition for it.  The basic local example is the behavior near a
conifold point, at which a conjugate pair of periods behaves as
$$
\Pi_A = z; \qquad \Pi_B = {\rm const} + {1\over 2\pi i} z \log z + \ldots .
$$
The corresponding flux superpotential is dual to the
$\CN=1$ SYM instanton superpotential \vafadual\ (we discuss this further
below); in this context it was studied in \gkp.
To get some global picture, one can consider the $T^6$ example, 
for which the periods are simply polynomial
in the complex structure $Z_{ij}$, already displays a lot of
structure, and we recommend that the reader unfamiliar with \CY3
look at \refs{\kst,\trivedi} as a start.

Let us pause to make some trivial mathematical remarks, which we find
important to get the right intuition about the configuration space
$\CC$ and this superpotential, which is rather different from what
intuitions based on branes or gauge theory on a compact space would
suggest.  What we have to say can be summarized mathematically as
follows: \CY3 complex structure moduli space, and (we conjecture) the
``true'' configuration space $\CC$ with stringy and quantum
corrections taken into account, is a hyperbolic space \kobayashi.

To illustrate what this means, we consider the simplest
``Calabi-Yau,'' the elliptic curve $T^2$ (equivalently, we could
discuss the dilaton-axion dependence in the \IIb\ problem).  Let its
complex structure is $\tau$, then its periods are $m + n\tau$ for
$m,n\in\BZ$.  They are not single-valued on the moduli space of
complex structures, which is the fundamental region in the upper half
plane, and thus the flux superpotential is not single-valued on moduli
space either.  

Rather, the periods are single valued on the Teichm\"uller space (by
definition), the upper half plane $\Im\tau>0$.  This is an open
complex manifold which (for purposes of studying these periods) cannot
be compactified.  Physically, this is to say that since a non-zero flux
breaks $SL(2,\BZ)$, there is no longer a unique large complex structure
limit, but rather many such limits.

In this situation, the ``topological'' counting formulas we discussed
earlier are not literally topological; by varying $W$ one can move
critical points $DW=0$ from the upper to the lower half plane.  Thus,
there is no immediate estimate of the form ``$N_{vac}=\chi(\CC)$'' for
numbers of flux vacua.  Related to this, a given period can take
values in a subset of $\BC$, and this is a possible behavior for a
superpotential on $\CC$.  All this is known for $T^{2n}$ and to some
extent for \CY3 moduli spaces, and we suspect it is the general
picture.

Returning to more local considerations, for generic fluxes, \gvww\
leads to a sufficiently complicated potential to make it very
plausible that critical points are isolated in all the variables it
depends on, the complex structure and axion-dilaton.  Although one
might worry that these critical points might be unstable to run away
to weak coupling (large $\Im \tau$) or large volume $\Im Z$, for
supersymmetric vacua this is not possible, as they necessarily have
$\Lambda\le 0$, while these limits have $\Lambda\rightarrow 0$.

Supersymmetric Minkowski vacua have been shown to exist; their
physical properties are discussed in \refs{\kst,\trivedi} and
many other works.
However, as it stands, this effective potential has no supersymmetric AdS
minima.  This is because $\p W/\p \rho=0$ and the special form of \Krho,
which forces
$$
g^{\rho\bar\rho} D_\rho W D_{\bar\rho} W^* = 3|W|^2
$$
and thus $V_{flux 2}\ge 0$, with equality if $D_i W=0$ in the
other moduli.  Since $D_i W=0$ is as many equations as unknowns, it
will have solutions; indeed this is just the problem of counting
critical points where we forget about the K\"ahler moduli.

This form for the potential is called ``no-scale structure''  and follows
because this is just another way to write
\fluxpotential, with its proper dependence on moduli
computed via KK reduction, and \fluxpotential\ is a positive sum of
squares.  The independence of $W$ on $\rho$ (which is exact in
perturbation theory) also implies that this potential does not
stabilize the overall volume (with more K\"ahler moduli, typically
some but not all are stabilized).

\subsec{Violation of no-scale structure, and the origin of $\Lambda_0$}

No-scale structure is a feature of the large volume, weak coupling
limit.  In any real model, further corrections will spoil this
structure and stabilize the K\"ahler modulus.  In fact, a ``no scale''
nonsupersymmetric vacuum with $V=0$, if it exists in the real theory,
will actually be a supersymmetric AdS vacuum.

Arguments have been made for explicit $\rho$-dependent corrections both
to $K$ and to $W$.  In \beckers, $\alpha'^4$ corrections to the 
ten-dimensional \IIb\ supergravity action were shown to produce the
following correction to \Krho,
$$
\delta K_{\tau,\rho} = {e^{-3\Im\tau/2}\over(\Im\rho)^3} + \ldots
$$
In \linde, it was recalled that 
nonperturbative effects in a $U(N_c)$ 
gauge theory sector will generically lead to exponentially
small corrections
\eqn\expW{
\delta W = M^3 e^{2\pi i\rho/N_c} + \ldots,
}
and there are stringy nonperturbative corrections of this form as well.

Either or both corrections spoil the no-scale structure.  A similar
correction spoiling no-scale structure can be found in $G_2$
compactification, by turning on a gauge field on an ADE singularity
supported on a hyperbolic 3-manifold, leading to stable AdS minima
\achflux.  In \linde\ it was shown that the correction \expW\ will
also lead to stable AdS minima, and that further effects can lift this
to a dS minimum, in a controlled regime (weak coupling and moderately
large volume).  

We regard these results as valuable evidence for our basic assumption,
that effective field theory can describe the physics of string/M
theory vacua.  However, we will have to make different arguments
to claim that this stabilization works at arbitrary coupling and
volume.

In the exact theory, one expects the K\"ahler moduli to be stabilized,
simply because of genericity.  In fact, the real problem is that they
are overdetermined; not only do we expect superpotential dependence as
in \expW, but if we had gone more deeply into the D-flatness
conditions in the brane sector, we would have seen that these already
stabilize one (real) K\"ahler modulus for each homology class of
simple brane in the construction (this type of argument can be found
at the end of \aspdoug), and a real treatment of this problem must bring
in the D-flatness conditions.  

However, to get some idea of the possibilities, let us simply assume
that nonperturbative physics produces a
superpotential which is the sum of \gvww\ and a correction
depending only on $\rho$,
\eqn\simplecor{
W = W_{flux} + f(\rho) ,
}
We retain \Krho\ for the K\"ahler potential.  This choice was made
because small corrections to $W$ can easily change the problem
qualitatively (by changing the $W=0$ locus), while small corrections
to $K$ generally do not.

In this case, the supersymmetry conditions for the complex structure
moduli and axion-dilaton are unaffected, while the $D_\rho$ condition 
becomes
\eqn\solverho{
0 = D_\rho W = D_\rho f(\rho) - {3\over\Im\rho} W_{flux}(Z,\tau) .
}
This equation determines $\rho$ in terms of $W_{flux}$ at a
critical point.  The resulting vacuum energy is
\eqn\newvac{
V = - 3 e^K |W_{flux} + f(\rho)|^2 ,
}
again implicitly a function of $W_{flux}$.

Now, the problem of finding critical points of $W_{flux}$ does not
have any obvious preferred scale, and it seems likely that by varying
the fluxes one can find critical points with arbitrarily large
magnitude $|W_{flux}|$, even taking into account the
anomaly cancellation condition \anomalyflux\ (since this is an
indefinite form).
In light of \newvac, this potentially violates our finiteness
conjecture.  There are two ways it could be saved, either by
cancellations in \newvac\ or by the possibility that if
$|W_{flux}|$ exceeds some upper bound $W_{max}$, the equation
\solverho\ will fail to have solutions.  The first possibility
requires an implausible conspiracy between $W_{flux}$ and $f(\rho)$,
so we consider the second.
This solution to the problem simply requires that the function
\eqn\fbound{
(\Im\rho) D_\rho f(\rho)
}
have an upper bound, which will be $W_{max}$.  This requires $f(\rho)$
to fall off as $\Im\rho\rightarrow\infty$, but we already know this
is true.  Requiring boundedness elsewhere more or less amounts to
requiring that $|f(\rho)|$ itself is bounded.  As we discussed, this
is quite possible; indeed the function \expW\ on the upper half plane
is an example.

It seems to us that some structure of this type is required to get
the number of flux vacua to be finite.  If we grant the simple
form of the equations \solverho, then we would conclude that the most
useful way to count supersymmetric flux vacua is to impose a bound
\eqn\Wbound{
e^K |W_{flux}|^2 < |W_{max}|^2
}
for some $W_{max}$, probably depending on the particular \CY3 and
other features of the compactification, and that the number so
defined should be finite. 

This would then lead to a lower bound on $\Lambda_0$, which in this
language unfortunately depends on nonperturbative physics.  It
might be that the origin of $\Lambda_0$ and any bounds it must satisfy
would be clearer in some dual picture.  Anyways, we offer this as
an argument for stabilization which could hold in general.

Unlike our other such arguments, the upshot of this one was not that
we claim K\"ahler moduli are stabilized for an $O(1)$ fraction of the
flux vacua.  Rather, we needed to call upon properties of some
(assumed) K\"ahler stabilization to even formulate the question of
counting flux vacua.  The formulation we end up with is the one given
in \fluxconj.

\subsec{The cross-coupling problem}

The flux potential is rather complicated but at least explicit, so we
can answer some interesting questions with it.  One of these is
whether in different vacua which might superficially agree with the
Standard Model, and have acceptably small cosmological constant, the
couplings are equal or at least similar, or whether they vary wildly
upon varying the fluxes.  On general grounds, Banks, Dine and Motl
\bdm\ suggested that the latter would be true, and more recently
this has also been pointed out by Acharya \acharyanew.

For example, let us consider a class of models which all contain a
common subsector of the model in which the Standard Model degrees of
freedom live.  For example, one can propose a configuration of branes
wrapped on cycles which realize the Standard Model matter content, and
which might be embedded in many different \CY3's.  This idea is
sometimes called ``modularity'' and is certainly natural from an
engineering point of view.  However, the question we come to is how
much of the structure of the rest of the
\CY3 we need to know about to predict any couplings.

We model the
situation by proposing two ``sectors,'' the ``Standard Model''
and everything else.  We have Standard Model fields $\psi$,
fields $Z$ which directly
control Standard Model couplings, and many more fields $y$ which
do not.  We then postulate a superpotential of the form
$$
W = \sum N_i \Pi^i(y,z) + W_{SM}(z,\psi)
$$
Suppose we vary the fluxes $N_i$ by an allowed quantized amount 
$\Delta N$ and find a new minimum; how much
do we expect the Standard Model couplings to change?  To simplify
the problem, we consider an infinitesminal variation $\delta N$;
although this is not physical, if $W$ and $K$ are not too rapidly
varying (which is generically true) $dZ/dN \Delta N$ will be
approximately the same.

One's first picture is that some $N_\alpha$ are associated to the ``cycles''
which we use to build the Standard Model, while most are not, and that
varying the $N_i$ which are not will make tiny corrections to the
$Z$'s.  This can be made more precise by computing $\p Z^i/\p N^j$
along the minima $W'=0$.  The only general topological relation between
cycles is expressed in the intersection form, $\eta_{ij}$.  Thus,
the type of general decoupling one might have expected would be true
if
$$
{\p Z^i \over \p N^j} \sim A_{ij} \delta_{ij} + B_{ij} \eta_{ij}
 + {\rm small\ corrections} .
$$

What actually happens if we vary satisfying $\delta (N^j D_k W_j) = 0$, is
that $\delta z$ determined by
$$
\delta N^j D_k W_j + N^j \delta z^i D_k \p_i W = 0
$$
can be large if $DW$ is large (in the other flux directions) or
$D^2W$ is small.  Now $D_k W_j \sim \tau_{kj}$ the matrix of 
$U(1)$ couplings.  At general points in moduli space,
these will not line up with $\delta_{ij}$.

This ``cross-coupling'' makes it difficult to claim that quantitative
aspects of one sector of the theory, such as couplings, can be
independent from quantitative aspects of another.
This is potentially another severe problem for the predictability of
the theory, and cannot be ignored.  However, it is not clear how much
of a problem it is, without having some real numbers.

\subsec{Brane-flux duality}

So far we discussed brane and flux degrees of freedom separately, but
it is known that this is overcounting, as many configurations have
dual descriptions of both types.  The prototype for this is of course
the seminal work of Maldacena \malda, which has been reinterpreted and
generalized in many ways.  For the purposes of string
compactification, perhaps the most useful form of this duality is the
``geometric duality'' of Gopakumar and Vafa \gv.

The simplest version of this states that a theory of $N$ D5-branes
wrapped on a small $S^2$ in a \CY3, which leads to $\CN=1$ $U(N)$
super Yang-Mills theory and a quantum generated superpotential, is
equivalent in a configuration with flux $N$ on a related \CY3 obtained
by replacing the $S^2$ with an $S^3$ (the ``conifold transition'').
This was proved (in a sense) in \ov, and many generalizations of the
basic result are known to more complicated geometries.

The full extent of brane-flux dualities is not known and we will have
to make a plausible guess to deal with this.  The most naive guess
would be that all branes can be dualized to flux, but this is not
possible as flux theories cannot have low energy non-abelian gauge
symmetry, while supersymmetric gauge theories with sufficient matter
(for example, $U(N_c)$ theory with $N_f>N_c$ flavors of matter) can.
It is amusing that the supersymmetric Standard Model provides 
an example, though because supersymmetry must be broken at a higher
scale than the strong coupling gauge scale it is not guaranteed that
this has deep significance.  In any case, we cannot realize the
Standard Model purely within the \IIb\ closed string sector.

A more sensible claim would be that brane theories which generate
superpotentials at the quantum level (for example, $U(N_c)$ SYM with
$N_f<N_c$ flavors of fundamental matter) can be dualized, while others
can not.  If so, then large numbers of brane configurations on \CY3's
are in fact redundant descriptions of the flux vacua, and should not
be counted.

A test of this idea, would be to compare the numbers of supersymmetric
vacua in a brane configuration on \CY3 $X$, with the number of flux
vacua in the geometric dual $Y$.  Since there are so many more brane
configurations than \CY3's, it is quite likely that many gauge theories
are in fact dual to the same \CY3 with fluxes, presumably to different
vacua within the latter theory.  Some field theory counterparts of this
possibility have been observed in \csw.  Given such an identification,
reproducing the same counting of vacua on both sides would be impressive
evidence for the duality.

The minimal test of this is that the number of flux vacua in the dual
theory should have $c^N$ multiplicity (in the new sector) as we argued
was generic for gauge theory.  Now the dual of a theory with $N$
distinct types of (B type) brane is a \CY3 with $2N$ new classes in
$H_3(Y,\BZ)$.  We have given arguments and will give more below that
this is the form we expect for multiplicities of flux vacua.

\subsec{To be continued}

At this point we have introduced more or less all the the ingredients
we will use to ``count vacua'' in section 5.  Although we somewhat
oversimplified them, we cannot do much better without a better
understanding of the many open issues we mentioned (and no doubt those
we didn't mention).  Furthermore, our picture has been too sketchy on
points such as orientifolding and anomaly cancellation, and has simply
left out a great deal, such as $E_8$ gauge symmetry (not visible at
weak coupling in \IIb) and other nonperturbative light states, the
brane world-volume superpotential, the detailed structure of stringy
nonperturbative effects, and so on.

Nevertheless, let us conclude with a final summary of how we will
combine the choices we just discussed in section 5.  Of course on a basic
level one picks a \CY3, a brane and flux configuration and so on; but to what
extent do these choices correspond to the more precise definitions of
vacuum counting we gave in subsection 3.2 ?

In $\CN=1$ \IIb\ theory, the chiral
multiplets can be divided into the complex moduli $Z$, the
dilaton-axion $\tau$, the K\"ahler moduli $\rho$, and the open string
modes $\psi$.  Techniques exist for computing classical brane
world-volume theories, with a superpotential $W_{cl}(\psi,Z)$ and
gauge couplings depending on $\rho$.  The D-flatness conditions
contain FI terms which also depend on $\rho$.  In fact many $U(1)$'s
are anomalous and these couplings are partners to the anomaly cancelling
couplings, lifting some of the $\rho$'s.

In general, we can expect gauge theory sectors with small matter
content (so, not including the Standard Model sector) to generate a
quantum superpotential stabilizing all their fields $\psi$.  On the
other hand, preserving supersymmetry in the Standard Model and other
sectors with large amounts of matter will tend to fix the moduli $\rho$
controlling their brane tensions (so that the different branes preserve
the same $\CN=1$ supersymmetry).

The simplest description of this physics would be to employ our conjectural
brane-flux duality to turn all of the branes for which quantum effects
lift the moduli space, into fluxes.  Thus, we will count brane configurations
with moduli spaces, multiplied by flux configurations, on each relevant
\CY3, and justify this by appealing to \sgindexprod.

Unfortunately, we are not keeping enough information in our considerations
to decide which brane theories have quantum moduli spaces; this depends
on the superpotential.  We will simplistically assume that all of the
hidden sector theories can have sufficiently complicated superpotentials
to produce isolated vacua and make this duality appropriate.  
Although this is clearly not always true (for example if we realized
another copy of the Standard Model, or a different fixed point theory),
it is plausible to claim that these sectors dominate the vacuum 
multiplicity.

While these assumptions are clearly an oversimplification, we will
still see in section 5 that interesting points will emerge from the
discussion.

\subsec{The main point}

To seriously address any questions of string phenomenology, we need to
make a discussion such as the one we just made, which exhibits the
various choices in string theory compactification, and derives the
consequences of each choice for the resulting four dimensional
effective theory.

Our discussion was terribly long and technical, to the point where it
is very hard to get any picture of how many possibilities will come
out, and how they are distributed.  And it skipped many important
points; a comprehensive discussion of these models would be far longer.

This is just how string theory is at present, and string theorists
must do the work to exhibit the potentially relevant vacua, to have
any solid foundation for string phenomenology.  Furthermore, since
most of the choices and consistency conditions have little direct
relation to the phenomenological considerations, it is difficult to
see how to do this without listing all or at least a representative
subset of the vacua.

But if listing the vacua produces an answer which is too 
complicated to think about, and the description of the procedure which
leads to the list of vacua is too complicated to think about,
then what can we do ?

\newsec{Ensembles of effective Lagrangians}

As one changes perspective from the problem of ``finding the right
vacuum'' to characterizing all the vacua, one realizes that the idea of
``list'' in many respects gives far more information than we actually
want, and is inflexible in a way which makes progress difficult.  A
more flexible concept might be an ``ensemble'' of vacua which assigns
a weight to each vacuum.  Armed with this concept, we might try, for
example, to find a simple ensemble which approximates the true ensemble
of superstring vacua well enough to address the goals we stated
in the introduction.

Now a vacuum, for present purposes, is a critical point of an
effective potential.  Most of its structure comes from where this
critical point sits in the effective theory, in the sense that small
fluctuations around it govern the spectrum and interactions of
particles.  Since we need to keep so much structure of effective
theory to say anything useful, we may as well change our concept to
instead define an ensemble of effective theories.  In words, we take a
subset of the data we need to specify the effective theory: field
content, potential, and other terms in the Lagrangian, and make this
particular data our precise definition of ``theory''; we then specify
an ensemble by giving a weight function on the space of these
theories.  A given theory could contain any number of vacua, and the
resulting ensemble of vacua is the sum of all of these vacua, each
weighed by the weight of the effective field theory which includes it.

The main examples we will discuss are ensembles of
$\CN=1$ supergravity theories with chiral multiplets and no gauge
multiplets.  The effective potential is then determined by a choice
of configuration space $\CC$, K\"ahler potential $K$ and
superpotential $W$.  These satisfy the usual rules of supergravity
\wessbagger: in particular, the K\"ahler form $\omega=\p\pb K$
is positive definite (since it is the kinetic term for scalar fields),
and the superpotential is a section of a line
bundle $\CL$ over $\CC$ such that $c_1(\CL)=-\omega$.

Physically, we will think of our supergravity Lagrangian as
a possible effective Lagrangian which might arise from some more
fundamental theory (e.g. string/M theory), defined in the usual
Wilsonian sense.
First, we have implicitly chosen an energy scale $M$.  All quantum
effects of virtual states with energies $E>M$ are included in
the effective Lagrangian.  On the other hand, the Lagrangian contains
all fields required to describe all particles with mass $m\le M$
in every vacuum.  It may also contain fields with $m>M$.

There is a lot to say about this dependence on scale and the role
of the renormalization group in these problems, and this has
been discussed in the phenomenology literature.  However,
for what we try to make precise in this paper, namely problems
involving counting of vacua and rough estimates of likelihoods to
match couplings, we do not need such a precise definition, and
can think of $M$ as infinitesimal.
In this case, a vacuum is defined as a critical point of the effective
potential, ${\p V/\p \phi^i} = 0$.

Thus, a ``theory'' for us is a triple $(\CC,K,W)$, and the set of
theories as the set of triples up to the usual geometric
identifications (field redefinitions).  This set has components in
each of which $\CC$ has a definite dimension, topology and complex
structure.  Each component is an infinite dimensional manifold,
a point of which is a choice of $K$ and $W$.

One can define natural metrics and even measures on these infinite
dimensional manifolds.  We will not need to go far into the
mathematics of this for the simple examples we give.  In any case,
given the ability to make such definitions, we can specify an ensemble
of theories by giving an integrable measure on the set of theories.
We will not require that it is unit normalized.

Let us give a simple example, to make this concrete.  Our example will
bear an obvious resemblance to the Gaussian ensembles of random matrix
theory.  It is a particular case of a class of ensembles studied as
models of quantum chaos, as we discuss below, and of course has
similarities to models commonly studied in the physics of disordered
systems such as spin glasses and random potential models.  Many have
suggested this general analogy (e.g. see \houches), and indeed spin
glasses share the $c^N$ multiplicity of vacua we observed in the
previous discussion \spinglass.  Below, we will add to these points of
similarity, the new observation that the class of ensemble we discuss
can be obtained by a simple limit of the flux superpotential.

We choose
$\CC$ out of the possibilities $\BC$, $\BC^2$ and so on, of arbitrary
dimension $n\ge 1$, and call the $n$ chiral superfields $z_i$.  Given a
particular choice for $n$ and thus $\CC$, we choose the K\"ahler
potential
$$
K = \sum_{i=1}^n |z_i|^2 .
$$
and superpotential
$$
W = \sum_{\vec I} w_{\vec I} z^{\vec I} ,
$$
a polynomial of degree $d$, where $z^{\vec I}=z_1^{I_1}\cdots z_n^{I_n}$.
Such a polynomial has $c(d,n)=(d+n)!/d!n!$ independent coefficients 
(the number of degree $d$ homogeneous polynomials in $n+1$ variables).

One ensemble (a Gaussian unitary ensemble or
``GUE'', since it respects $U(n)$ symmetry)
could be defined by taking the coefficients 
$w_{\vec I}$ to be complex.  We could define a different ``GOE''
ensemble by taking the coefficients real, which would be appropriate if the
systems of interest had CP symmetry.
In either case, we choose the coefficients with weight
$$
[d\mu(W)] = \left({\pi\over\alpha}\right)^{c(d,n)}
 [\prod_{\vec I} dw_{\vec I}] e^{-\alpha \sum_{\vec I} |w_{\vec I}|^2} .
$$
(this is unit normalized).

Finally, we can specify a weight $P_n$ for each of the possible
dimensions $n$.  If we were introducing randomness purely as a
theoretical device to model generic superpotentials, we should take
$P_n=\delta_{n,N}$, {\it i.e.} simply fix $n$ and unit normalize.  
On the other hand, if
the randomness reflects a multiplicity of effective field
theories arising from a more fundamental theory, we should choose
the $P_n$ to reflect these multiplicities.

Clearly this simple example can be generalized in many ways, and
we will discuss a few aspects of this below.
One can also generalize
it to add gauge symmetry.  One would have a weight $P(G)$ for
each particular choice of gauge group $G$, and randomly 
chosen actions by isometry of the gauge group on $\CC$ (in the
simple example of flat $K$ this is the choice of matter representation),
of Fayet-Iliopoulos parameters, and of more exotic couplings.

We define the expected number
of vacua, meaning critical points of $V$, in the ensemble, as
\eqn\Nvac{
\vev{N_{vac}} = \sum_\CC P(\CC)
 \int [d\mu(K,W)]\ \int_\CC\ [d^{2n}z] \delta^{(2n)}(V')~ |\det V''| .
}
The determinant is present in order to count each isolated vacuum with
weight $1$, and give a coordinate-independent result.
There are some variations one can make on this which we will discuss
shortly.

The first point we want to make is simply that given an ensemble,
one could compute mathematically or estimate on a computer the
number $\vev{N_{vac}}$.  
We will discuss some results of this type
in an extremely simplified example below, and in more examples elsewhere.

Indeed one could in principle compute
the expectation value of any physical observable in a given 
theory in the ensemble, by integrating the observable over the
ensemble.  For example, one could compute the distribution of
cosmological constants,
\eqn\cosdist{
\rho(\Lambda) = {1\over N_{vac}}\sum_\CC P(\CC)
 \int [d\mu(K,W)]\ \int_\CC\ [d^{2n}z] \delta^{(2n)}(V')~ |\det V''| ~
 \delta(\Lambda-V) .
}

The second point is that some results need not depend on the details
of the ensemble.  As the simplest example of this, since the number of
critical points of a generic function is invariant under small
perturbations of the function, the number of vacua will remain invariant
if we ``fuzz out'' the ensemble, replacing delta function contributions
to the measure with highly peaked functions of unit weight.  This
allows a lot of potential scope for simplifying the ensemble.

One can also entertain the hypothesis that a sufficiently complicated
potential will start to look like a generic member of a simple
ensemble.  This is expressed more precisely in the idea of
``universality,'' which we will return to.

Let us define some related observables which are more or less precise
than \Nvac.  First, one can discuss the number of perturbatively
stable vacua, meaning those without tachyons.
In Minkowski space (zero cosmological constant), this means of course
that the matrix $V''$ must be positive definite (it is related to the
mass matrix by the positive definite metric on field space).  Thus,
we define
\eqn\Nstablevac{
\vev{N_{stablevac}} = \sum_\CC P(\CC)
 \int [d\mu(K,W)]\ \int_\CC\ [d^{2n}z] \delta^{(2n)}(V')~ |\det V''|
~ \theta(V'')
}
where the $\theta$ function applies to each eigenvalue of $V''$.

One can go on to make the question more specific by imposing
metastability with respect to tunneling (obviously this is harder
to treat analytically, but some useful tricks appear in the theory
of spin glasses \spinglass), or ask for vacua with certain qualitative
properties.

One can also consider less specific questions of this type.  For
example, the integrand in \Nvac\ is not the simplest one could consider.
A simpler possibility is
\eqn\Nindex{
\vev{I_{vac}} = \sum_\CC P(\CC)
 \int [d\mu(K,W)]\ \int_\CC\ [dz] \delta^{(2n)}(V')~ \det V'' .
}
where we do not take absolute value of the determinant, and thus
count vacua with signs $\pm 1$ depending on its sign.  This type of
signed measure is familiar in supersymmetry, 
topological field theory, and so on,
and produces the supergravity index \sgindex.
As such, it should be much easier
to compute, yet also gives information which might be useful in
understanding $N_{vac}$ or even its more specific relatives.

\subsec{Estimated number of nonsupersymmetric vacua}

To illustrate how very simple estimates of numbers of vacua can be
made in this language, we consider an ensemble of globally
supersymmetric theories with a Gaussian distributed superpotential,
and explain how to get estimates for the density of supersymmetric
vacua and nonsupersymmetric vacua.

We start with the ensemble of theories with $n$ chiral superfields $z^i$
taking values in $\BC^n$, and a superpotential
\eqn\degreethree{
W = w + a_i z^i + b_{ij} z^i z^j + c_{ijk} z^i z^j z^k .
}
The $(n+1)(n+2)(n+3)/6$ coefficients are taken from independent distributions
with the Gaussian weight
\eqn\gaussone{
d\mu[W;\alpha] = [dw da db dc] N(\alpha) \exp -\alpha 
 \left( |w|^2 + \sum_i |a_i|^2 + \sum_{i\le j} |b_{ij}|^2 + \sum |c|^2 \right)
}
where $N(\alpha)$ is chosen to normalize the total weight,
$$
1 = \int d\mu[W;\alpha] .
$$
This leads to
$$
N(\alpha) = \left({\alpha\over\pi}\right)^{(n+1)(n+2)(n+3)/6} .
$$

Note that this ensemble is symmetric under $U(N)$ rotations
$z\rightarrow Uz$, but it is not translation invariant; taking
$z\rightarrow z+z^i$ changes it.  In particular, the lower moments of
$W$ have a nonzero expectation value in the shifted fields.
One can find ensembles with more symmetry (say $U(n+1)$)
at the cost of changing the K\"ahler potential, and one might expect
these to play a more important role in describing more natural
starting points such as string/M theory.
In any case, this ensemble is fine for purposes of illustration.

We start by considering the expected number of supersymmetric vacua.
Of course we already know this for the superpotential \degreethree.
We need to solve the $n$ equations $W'=0$.  Each of these has degree
two, and a generic system of $n$ degree two equations in $n$ unknowns
will have $2^n$ solutions.

Let us instead do this using the formula
\eqn\Nsusyvac{
\vev{N_{susy}} = 
 \int [d\mu(W)]\ \int_\CC\ [dz d\bz]~ \delta^{(n)}(DW)~ \delta^{(n)}(\bar DW) 
~|\det D^2W|^2 .
}
Since we are discussing global supersymmetry, we take $D_i=\p/\p z_i$ the
usual coordinate derivative.  It is clear at this point that the constant
factor $w$ will drop out of our considerations, so we now set $w=0$
(and remove the corresponding factor from $N(\alpha)$).

One advantage of this approach is that we can get not just a total number
but an expected density of supersymmetric vacua $d\mu_{susy}(z)$, defined
by
\eqn\Nsusyvac{
d\mu_{susy}(z) = [dz d\bz]  \int [d\mu(W)]\ 
\delta^{(n)}(DW(z)) \delta^{(n)}(\bar DW(\bz)) 
|\det D^2W(z)|^2 .
}
At a fixed $z$,
this is a simple Gaussian integral.  The integrand depends on $z$, but
in a relatively simple way.  

One can explicitly compute the density as a function
of $z$.  To make our point, we will just do it for $z=0$.

At $z=0$, we have $D_iW(0) = a_i$ and $D_i D_j W(0) = b_{ij}$.
The delta functions can simply be used to set $a_i=0$, while the
values of $c$ now drop out of the discussion.  This leads to
\eqn\Nsusyvaczero{
d\mu_{susy}(0) =
\left({\alpha\over\pi}\right)^{n(n+3)/2} 
[dz d\bz] 
\int [db] e^{-\alpha \sum_{i\le j} |b_{ij}|^2}
|\det_{i,j} b_{ij}|^2
}
This Gaussian integral is not hard to do; the result is
the probability that a randomly chosen superpotential will have a 
supersymmetric vacuum at the origin, or equivalently the expected
number of vacua at the origin in this ensemble.
It is not much harder to compute the density $d\mu(z)$,
which turns out to have power law falloff.

Let us compare this to the computation
of the expected density of all vacua $V'=0$, including
nonsupersymmetric vacua, evaluated at $z=0$.  Starting with
\Nvac\ with $V=\sum_i |D_iW|^2$ and following the same approach leads to
\eqn\Nallvaczero{\eqalign{
d\mu_{vac}(0) =
\left({\alpha\over\pi}\right)^{(n+1)(n+2)(n+3)/6-1} 
&[dz d\bz] \cr
\int [da\ db\ dc] e^{-\alpha \sum |a|^2 + \sum |b|^2 + \sum |c|^2} 
&\delta^{(n)}(b_{ij} a^*_j) \delta^{(n)}(b^*_{ij} a_j)
|\det M|
}}
where $M$ is the matrix of second derivatives $V''$, a
$2n\times 2n$ hermitian matrix
$$
M_{i,j} = \left(\matrix{
b_{ik}b^*_{kj}& c_{ijk} a^*_k \cr
c^*_{ijk} a_k & b_{ik}b^*_{kj} }\right) .
$$
One then solves the $\delta$ function constraints for $b$,
since in a nonsupersymmetric vacuum $a\ne 0$, and proceeds above
to obtain a density,
which could be integrated to obtain the total expected number
of nonsupersymmetric critical points.  Such results and the
more interesting expected number of stable nonsupersymmetric vacua
will be discussed in \refs{\dougzel,\dougashok}.

\subsec{Universal results}

The first question one should ask about the type of result we just
described is to what extent they are particular to a specific
ensemble, and to what extent they reflect properties shared by many
ensembles and which it is reasonable to believe are shared by the
ensemble of string/M theory vacua.

In general, results do depend on the specific choice of ensemble.  In
\gaussone, this includes the choice of equal variances for $a$, $b$
and $c$; clearly the choice of distribution which weighs discrete
factors such as gauge group and matter content will be even more
important.  Claiming that a specific choice reproduces some aspect of
the ensemble of string/M theory vacua is therefore non-trivial.  Thus,
a good answer to this question requires some ability to work with the
string/M theory ensemble, and this is why the simpler considerations
of this section would have little content without the more complex
discussion of section 3.

However, the most interesting and potentially useful questions would
be those whose answers display universality.  There are various ways
one can try to make this concept precise, but one typical and standard
notion is to consider a family of ensembles parameterized by an
integer $N$, and claim that the $N\rightarrow \infty$ limit is a
universal function of a few parameters which can be extracted from the
ensemble, while $1/N$ corrections to the leading behavior may be less
universal.

In work of Bleher, Shiffman, and Zelditch on quantum chaos
(\refs{\zelditch,\zelreview}, and see also the references there); such
a universal limit is discussed for essentially the ensemble we just
discussed.  One considers a compact K\"ahler configuration space
$\CC$, and a positive line bundle $\CL$.  One then considers a family
of Gaussian distributed holomorphic sections $W_N\in \CL^N$ with the
measure (here $({\rm vol_\omega})=\omega^n/n!$ is the volume form):
$$
d\mu[W] = \exp -\int_\CC ({\rm vol_\omega})\ e^{N\cdot K} |W|^2 ,
$$
and considers average properties of the critical points of $W$.

As $N\rightarrow\infty$, using the simple estimates we discussed earlier,
the number of critical points will grow as $N^n$ (for $n$-dimensional
$\CC$), and thus one expects their average spacing to go as $N^{-1/2}$.
In the limit, the distribution of critical points becomes universal:
\eqn\expcrit{
d\mu[W] = {(\rm vol_\omega}) \times c_n~N^n~ (1 + \CO({1\over d})) ,
}
with $c_n$ a universal dimension dependent constant.

These results are essentially local and one might expect them to hold
for suitable sums over sections even in the case of more physical
interest, in which $\CC$ is noncompact.
This supports our earlier formula \sgindex\ as it gives a limit of
the problem in which this formula becomes exact.

If one focuses on the structure at the scale of the average spacing
between vacua by defining $Z = N^{-1/2} z$, one finds in addition that
correlation functions involving the products of densities of vacua at
distinct points become universal.  In this sense, the local structure
of the effective field theory becomes universal.  Physical
applications of such results might include computing the probability
that another nearby vacuum could destabilize the vacuum of interest,
or that flow from one critical point to another realize sufficient
inflation.  Such questions will be studied in \dougzel\ and future
work; these are questions which one has little hope of addressing at
present except in ensembles.

Finally, the parameters which entered into this example (derived from
the K\"ahler metric and parameters of the distribution) will control
the expected number of nonsupersymmetric vacua as well.  This is the
sense in which we would claim that general information about
supersymmetric vacua can determine numbers of nonsupersymmetric vacua,
which we will explore in \refs{\dougzel,\dougashok}.

\subsec{The ensemble of string/M theory vacua}

Besides explicitly defining ensembles, one can implicitly define
ensembles.  Obviously the primary one we are interested in would be an
``ensemble of all theories coming out of string/M theory.''  This
ensemble is not computable at present, but perhaps if we can precisely
define it, some approximation to it will someday be computable.

The basic idea one wants to implement is the following.  In section 3
we discussed the many, many choices which enter into the construction
of a compactification in string/M theory.  As is much discussed in the
literature, each of these choices leads to a low energy effective
Lagrangian, with specific field content and couplings, valid in a
certain region of configuration space.  We want to define an ensemble
for which the measure which is a sum of delta functions, one for each
effective field theory which is obtained from a given choice of the
discrete compactification parameters.

Each effective field theory in the ensemble so defined will in general
describe many vacua.  Although so far we have stressed the idea that
the ensemble will contain distinct effective field theories each
corresponding to one choice of the compactification data, we could
also imagine that a given compactification is not described by a
single effective theory but rather by a collection of ``dual
descriptions,'' each with partially overlapping regions of validity.
To do this well, we must generalize our definition of ensemble, for
the following reasons.

First, we may only trust a given effective field theory if the fields
live in a certain region of configuration space.  This is definitely
not a problem of principle and there is no {\it a priori} restriction
on the region which can be used.  In particular, there is no reason
that one effective field theory cannot be valid over a range of
configurations with relative distance large compared to $M$, or even
large in Planck units.  However, it often does happen that a specific
derivation breaks down in such circumstances (typically because new
light states come down), so one needs to allow for this type of
partial information.  Now there is a way to deal with it given our
previous definitions: we simply take $\CC$ to be a manifold with
boundary, which cuts out the regions we do not trust.  However, this
will lead to many complicated regions $\CC$ and a complicated
description.  We would prefer to describe this information more
simply.

Second, different effective field theories can be dual, and describe
the same physics.  While the observables that we compute should
reflect these identifications, this does not immediately bear on the
question of whether we count two dual vacua in dual effective
field theories as equivalent or not.  If they do not come from dual
underlying compactifications, they clearly count as distinct possible
vacua in any reasonable theoretical formulation.  Since we already
have weights, there is no difficulty in representing this.

Third, it may be that two choices of the discrete compactification
data are dual and lead to the same set of vacua.  It could
even be that two choices are only partially dual: that there is a
subset of vacua of theory A which are to be identified with a subset
of vacua in theory B, while the other vacua are distinct (or perhaps
identified with other theories).  This leads to problems of description
analogous to the first point above.

All these features lead to complexity in the relation between the starting
point (string/M theory) and the final ensemble, but they need not imply
inherent complexity in either the starting point or the final ensemble.
Nevertheless we need some formal language to describe it.

The suggestion we will make to treat it is the following.  We generalize
our weight $P_n$ which represented the weight of a given configuration
space $\CC$, to a weight function on $\CC$, a real function 
$\rho(\CC;\ z,\bz)$.
And we generalize the definition of the expected number of vacua to
$$
\vev{N_{vac}} = \sum_\CC
 \int [d\mu(K,W)]\ \int_\CC\  \delta^{(2n)}(V')~ |\det V''|~
 \rho(\CC;\ z,\bz) .
$$
This is a rather broad generalization which certainly allows the
freedom to deal with the problems we just discussed.  If we only
trust a given effective field theory to describe a subset of
configuration space, we set $\rho=0$ outside that subset.  If we
have two dual theories which each describe a region $R$, we set
$\rho=\half$ for each of the theories within $R$, and so on.

Obviously this is a highly redundant and ambiguous description, and it
would be implausible to claim that string/M theory leads to particular
preferred $\rho$'s, $\CC$'s, and so forth.  On the other hand, to the
extent that string/M theory leads to many different effective
theories, it is not {\it a priori} implausible to claim that the sum
of the contributions to the ensemble from many unrelated effective
theories produces a much simpler ensemble, with simple choices for
$\rho$ along with the rest of the data, than any of the individual
effective theories might suggest.  Such a claim should be evaluated by
comparing results computed in a given ensemble to results derived
directly from sets of many actual compactifications of string theory,
in a spirit somewhat analogous to the original tests of superstring
duality.

We are not claiming that all physical questions about string/M theory
can be usefully addressed this way.  But to answer some questions,
starting with the number of vacua, and going all the way to our
primary question of how many vacua well approximate the Standard
Model, we do not need to reproduce the stringy ensemble precisely.  To
the extent that string/M theory has finitely many vacua (with
``reasonableness conditions'' of the type we suggested in section 3),
the true ensemble will always look like a sum of delta functions, and
indeed this is where the predictive power of string theory lies.  On
the other hand, we might well get good estimates for the quantities we
have stressed as potentially accessible to this approach by using an
ensemble with a smooth measure on theory space, which could be far
simpler than the true ensemble coming from string theory.

We conclude by stressing the importance of the claim that this ``true
ensemble of string/M theory effective field theories'' exists and is
well defined, up to ambiguities in describing dual sets of theories by
specific choices of $\rho$.  The existence of this ensemble depends on
only two essential points.  First, we need a precise definition of
string/M theory, in which vacua are described by specific effective
field theories.  Second, we need ``local finiteness'' of the ensemble,
meaning that there should not be an infinite number of distinct
(non-dual) compactifications which lead to the same or arbitrarily
similar effective field theories.  One of the main points of the
discussion in section 3 was to give arguments that the number of
physically relevant vacua is finite, which would certainly imply local
finiteness.  

Granting that this ensemble exists, to some extent it and
approximations to it can play the role for $\CN=1$ and
nonsupersymmetric theories, which moduli spaces of vacua played for
studying theories with extended supersymmetry.  The subsequent
examples will illustrate this point.

\subsec{The flux superpotential ensemble}

We can define the ensemble of flux superpotential vacua as follows.
We take $K$ to be the standard K\"ahler metric.  And we take $W$
distributed as
$$
d\mu(W) = \sum_{N\in \BZ} \delta(W - N_i \Pi^i(z)) 
 e^{-\alpha_{ij} N^i N^j}
$$

The exponential factor could be used to get convergence of the sum,
and a finite number of vacua, and also to enforce the $\eta_{ij} N^i
N^j$ anomaly cancellation constraint, by coupling this to a parameter
and taking an integral transform.  The other constraints we discussed
in section 3
enter in different ways: the cosmological constant is an observable,
while excluding the large volume limit would require cutting this
region out of the complex structure moduli space, and exclusion of
the large K\"ahler modulus limit (in \IIb) would enter into the
discussion in subsection 3.9.

The relation between this ensemble and the Gaussian orthogonal
ensemble we discussed in the previous section should be clear: we get
the latter from the former by forgetting about the quantization
condition on the $N^i$.  In other words, we ``fuzz out'' the delta
functions.

The main point we take from this is that simple Gaussian ensembles
such as those studied by \zelditch, are actually quite similar to the
actual ensemble of flux superpotentials, lending support to the idea
that they can well approximate their physics.  We discuss this further
in \dougzel.  

One can take the same type of $d\rightarrow \infty$ limit we discussed
above, by considering $K\rightarrow d K$ and superpotentials of the form
$$
W = \sum_N N_{i_1\ldots i_d} \Pi^{i_1} \cdots \Pi^{i_d} .
$$
Since $K$ always enters as $K/M_{pl}^2$, the scaling
$K\rightarrow d K$ corresponds
physically to ``taking $M_{pl}\rightarrow 0$'' or in other words
considering a series of problems in which the effects of supergravity
(compared to global supersymmetry) become increasingly important.
In any case, the limit distribution is (plausibly) \expcrit;
the vacua become uniformly distributed with respect to the 
volume form derived from the  K\"ahler form.

Since the fields entering into this superpotential will control
coupling constants in the observable sector, this claim gives some
precise meaning to the idea that ``flux vacua are uniformly
distributed in the space of couplings.''  
Uniform here means with respect to the K\"ahler metric on moduli
space.

\subsec{Uniform ensembles of effective $\CN=1$ gauge theories}

Here we discuss simple ensembles of low energy effective theories,
which are not directly motivated by string/M theory.  
These are important because they are the simplest possible guess as to
what will come out.  Much more thought should be given to what would
be useful here, taking into account the RG and the eventual
phenomenological tests. 

Let us just give two very simple examples.  
The first is a simple expression of the traditional idea of naturalness,
motivated by perturbative renormalization group considerations.
The second would be appropriate for a theory with a duality symmetry.

We fix the K\"ahler potential (say it is canonical), and fix a gauge
group $G$, and matter with a linear gauge action in representation
$R$.  We enumerate all gauge-invariant couplings $g_k z^k$, where
$g_k$ has canonical mass dimension $3-k$.

An ensemble is then specified by a distribution for these couplings.
Of course, we cannot simply integrate over all couplings with Lebesgue
measure, as this distribution is not normalizable.  

An obvious requirement to impose is that the couplings be natural
with respect to a UV scale $M$; in other words $g_k M^{k-3}$ should
be $O(1)$.  Let us also ask that dimensionless couplings are $O(1)$.

We choose a positive number $\alpha$, and take couplings with weight
$$
d\mu[W] = \exp -\alpha M^{6+n} \int_{B_M} |W(z)|^2\ d^nz ,
$$
where $B_M$ is the ball $\sum |z^i|^2 \le M^2$, with the volume form
derived from the K\"ahler metric.  This more or less says that the sum
of dimensionless couplings squared, measured in units of the cutoff,
is at most $1/\alpha$.

The Gaussian form is a choice of course; one could also bound the
integral of $|W|^2$ above, or try other weights.  It would be interesting
to decide which choice best reflects traditional ideas of naturalness.

One might also try to fix the arbitrary choice of $\alpha$ by taking
$\alpha=1$, motivated by the idea that a theory
with coupling $g>1$ should be dual to another theory with coupling
$1/g$.

To better implement this idea, let us define the 
$(G,\Gamma)$-uniform ensemble,
appropriate to a theory with duality symmetry.
Suppose we know a theory admits $\Gamma$ duality symmetry,
where $\Gamma$ is a discrete group.
Typically, the duality group $\Gamma$ is a discrete subgroup 
of some continuous group, $G$, with a natural action on the 
couplings.  In this case, the natural or ``$(G,\Gamma)$-uniform''
ensemble
(or, if there is an obvious candidate $G$,
the $\Gamma$-uniform ensemble),
uses a measure with
$G$ symmetry, and integrates over a fundamental region of $\Gamma$.

For example, the natural ensemble of $\CN=4$ supersymmetric
gauge theories according to this criterion is $SL(2,\BZ)$-uniform,
where the complex gauge couplings $\tau=\theta/2\pi + 8\pi i/g^2$
are distributed according to the measure
$d\tau/(\Im\tau)^2$, over a fundamental region of $SL(2,\BZ)$.

\subsec{Ensembles of quiver gauge theories}

The minimal information in an ensemble of gauge theories is a distribution
$d\mu(G,R)$ over the choice of gauge group and representation.
In the case of $U(N)$ quiver gauge theories, this data is very simple: the
gauge group is a product $\prod U(N_i)$ and thus specified by a list
of non-negative integers $N_i$ with $1\le i\le K$,
while the matter content is specified
by the $K \times K$ matrix $C_{ij}$
of multiplicities of chiral multiplets in the $(\bar N_i,N_j)$.
This will be the intersection matrix $I_{ij}$ of \intnum\ plus a
symmetric matrix if non-chiral matter is also present, which is
generically not expected.

Thus, an ensemble of quiver gauge theories will be specified by
a distribution $d\mu(N_i;I_{ij})$.  This could also depend
on additional information such as couplings, but we will not consider this.

We can now ask the question: if we consider the collection of all
\CY3's (that we know about) and all quiver gauge theories which arise
as world-volume theories of type \II\ branes, what ensemble do we get?  
Since we have techniques for computing these gauge theories, this
question could be studied by various means, for example by Monte Carlo
({\it i.e.} generating examples by computer).

We first consider a toy example which can be solved, and use it to
motivate some simple model ensembles which illustrate
some likely features of the true ensemble.  The toy example is the
ensemble of supersymmetric theories which describes bound states
of two elementary branes $B_1$ and $B_2$, with $k$ bifundamentals in
the $(\bar N_1,N_2)$.  As discussed in \refs{\dfr,\trieste},
the spectrum of bound states (simple objects) in this theory is easy to 
describe completely.  Assuming a positive FI term,
if and only if the dimension of the gauge group minus the number of
chiral matter fields is $1$, then 
there will be a unique supersymmetric configuration (up to gauge equivalence)
breaking to $U(1)$, i.e. a simple object.
Thus, there is a unique bound state $N_1 B_1 + N_2 B_2$
for each pair $(N_1,N_2)$ satisfying
$$
N_1^2 + N_2^2 - k N_1 N_2 = 1 .
$$
The net number of chiral multiplets between two such simple objects $B'$
and $B''$
with charges $(N'_1,N'_2)$ and $(N''_1,N''_2)$ is the intersection number,
$\vev{B',B''} = N'_1 N_2 - N'_2 N_1$.  Thus a supersymmetric brane theory
made up of such bound states is specified by a list of simple objects
with multiplicity; the spectrum of chiral multiplets is determined by
the intersection numbers, and the resulting theory will have no
superpotential.

Furthermore, this spectrum satisfies a symmetry under a simplified
form of Seiberg duality (the reflection functor), which acts as
$(N_1,N_2) \rightarrow (k N_1 - N_2, N_1)$.  In fact, the charges of
simple objects are real roots of the Kac-Moody algebra with
generalized Cartan matrix 
$\left(\matrix{2& -k\cr -k& 2}\right)$, and these transformations act
on the charge vectors as Weyl reflections.  All simple objects
can be obtained from one of them by a sequence of Weyl reflections,
and this leads to a set parameterized by a single integer $n$.  
Let $B_n$ with $n\le 0$ be the sequence $[B_0]=(k,1)$, 
$[B_{-1}]=(k^2-1,k)$, etc. and $B_n$ with $n>2$ be the sequence
$[B_3]=(1,k)$, etc.

The simplest resulting world-volume theories are those
describing combinations of two simple objects $B_a$ and $B_b$ (which
is enough to get all others as bound states).  
The Weyl reflection symmetry can be used
to take $B_a=B_1$; then if
$[B_b]=(N'_1,N'_2)$, the
intersection number will be $N'_2$, and the resulting collection of
quiver theories will be the set of $U(N_a)\times U(N_b)$ theories
with intersection number $I_{ab}=N'_2$.  Thus the corresponding
distribution is precisely
\eqn\exactIdist{
d\mu(I_{ab}) = \sum_{n\ne 1} \delta_{I_{ab},\vev{B_1,B_n}} .
}

Now the approximate growth of $N_2$ with $n$ is $N_2 \sim k^{|n|}$
so this distribution can be approximated as
\eqn\approxIdist{\eqalign{
d\mu(I) &= \sum_{n\ge 1} \delta_{I,k^n} + \delta_{I,-k^n} \cr
 &\sim dn (\delta(I-k^n) + \delta(I+k^n)) \cr
 &\sim d\left({\log |I|\over\log k}\right) \cr
 &\sim {1\over\log k}\left({dI\over |I|}\right) 
}}
where $I$ is a non-zero integer.

Now, if we have a fixed $k$ and a single sequence of world-volume
theories, since the actual distribution \exactIdist\ is 
sparse, one might not regard
this approximation as very accurate.  On the other hand, if
one was considering a large collection of such theories for varying $k$,
say arising by wrapping branes on different sets of cycles in different
CY's, one would add a variety of distributions of this form, and generically
produce a total distribution
$$
d\mu(I) \sim {dI\over |I|} .
$$
This illustrates the type of simplification we might gain by considering the
problem as a whole.

Note that there are an infinite number of theories, and the
distribution is not normalizable.  In the string theory application,
the distribution is finite only with
the tadpole cancellation constraint,
which will set the K theory class of the resulting configuration.
In the toy example, this could be modelled by setting $N_1$ and $N_2$,
and considering theories which can be obtained by partitioning these
fixed numbers among simple objects.  One could also obtain representative
results by placing a cutoff at $I\sim N_{max}\sim 100$, the generic
order of the Chern classes which enter this condition.  This suggests
the normalized distribution
$$
d\mu_{norm}(I) \sim {1\over 2\log N_{max}}\theta(N_{max}-|I|){dI\over |I|} 
$$
as the final approximate distribution for the toy model.

We now consider generalizing this to the case of $N$ nodes.  We could
proceed the same way in terms of the spectrum of bound states of $N$
elementary objects; quite a bit is known in simple cases and we will
return to this in subsequent work.  However the basic constraint which
was operating in the toy example was a simplified form of Seiberg
duality.  While there we used the action of
Seiberg duality on both the $N_i$ and $I_{ij}$, since
the ranks of gauge groups $N_i$ in the end were largely constrained by anomaly
cancellation, and the interesting result was for $I_{ij}$,
we might try to look for a distribution $d\mu(I_{ij})$
invariant under \seiint; in other words
\eqn\seiinv{
d\mu(I_{ij}) = d\mu(I_{ij} - 2I_{in} - 2I_{nj} + I_{in}|I_{nj}|)
}
for any choice of $n$ (of course a natural solution would also be
invariant under permutation on the indices and this choice would not
matter).  
This is a linear condition on the ensemble, because each application of
Seiberg duality relates one theory in the ensemble to another single
theory in the ensemble.\foot{The nonlinear
condition which appeared in the original version of the paper
would have been appropriate
for a transformation which combined different theories, but is incorrect
here.}

Thus, it is an interesting question whether there are ``natural''
distributions which satisfy \seiinv.  Note that the reason to think
that the correct distribution approximately
satisfies \seiinv\ is a bit subtle as it
is {\it not} because we are describing a set of theories which are
all related by Seiberg duality.  Rather, it is analogous to the idea
we discussed above that a randomly generated ensemble of (for example)
$\CN=4$ SYM theories should have couplings $\tau$ drawn from the 
$SL(2,\BZ)$-uniform
distribution.  After making duality identifications on the resulting 
space of theories, generic distributions which are generated by other
means will tend to map into the simplest distributions compatible with
the duality.

The simplest candidate dual distribution is of course the uniform
distribution with $d\mu(I)$ constant and independent of $I$.  While we
cannot claim to have ruled this out, it does not fit with putting
bounds on the numbers of branes (tadpole cancellation), the toy model,
or experience with brane construction (which tends not to produce models
with large intersection numbers), so we do not consider this the
preferred choice.  Rather, it seems more likely to us that the correct
distribution shares the general power law $dI/|I|$ behavior of the toy
model, which we will refer to as ``duality scaling'' behavior here.

To illustrate this within the simplest possible model ensemble, we can
take the components $I_{ij}$
with $i<j$ to be independently power law distributed, say as
\eqn\Idist{
d\mu(I) = \alpha \delta_{I,0}
 + \beta \sum_{n\ge 1} n^\gamma (\delta_{I,n} + \delta_{I,-n}) .
}
Within the independent component approximation, the condition 
\seiinv\ is roughly
$$
d\mu(|I|) \sim d\mu(|I|^2)
$$
where $|I|$ is the average magnitude of a component of $I_{ij}$.
Leaving aside the uniform distribution, this is best solved by 
$dI/|I| \sim d(I^2)/|I^2|$,
so that $\gamma=-1$ is preferred by this condition.

Thus, we adopt as our simple model of the ensemble of quiver gauge
theories arising from CY compactification, the distribution \Idist\ with
$\gamma=-1$, and a cutoff 
$$
|I| \le N_{max}
$$
as in our previous discussion, leading to the (approximate) normalization
condition
$$
1 = \alpha + 2\beta \log N_{max} .
$$
The coefficients should be $O(1)$ and we will simply take $\alpha=1/2$,
so that $\beta = 1/4\log N_{max} \sim 1/18$ for $N_{max} \sim 100$.

This leads to a unit normalized ensemble, representing the fraction of
quiver theories which are expected to have a particular chiral matter
spectrum.  This should be multiplied by the total number of quiver
theories to obtain an ensemble of theories.  In the toy model, this
number was $c=\log N_{max}/\log k$; more generally it seems plausible
that this number has the general form of our previous estimates,
perhaps $c^K$ or $c^{N_{max}}$.  As we discussed in section 3,
brane-flux dualities suggest that the total number of inequivalent
choices of brane and flux is comparable to the total number of choices
of flux, and we will make our estimate in section 5 under this
assumption.

While this ``duality scaling'' ensemble illustrates some properties of the
true ensemble of quiver gauge theories which come out of brane
constructions on \CY3, it is probably too simple to be very realistic.
In particular, the assumption of complete independence between intersection
numbers is probably false.  For example, one might think that a group
of branes wrapped on cycles obtained by resolving a single isolated
singularity, would be likely to have zero intersection numbers with
other branes.  

A second model ensemble which at least qualitatively reflects this
structure would be to take $K\times K$ intersection matrices which are
block diagonal and are direct sums of intersection matrices of
dimension $K_i$ distributed according to the previous ``duality scaling''
ensemble.  Such a direct sum is labelled by a partition of $K$ into
positive integers $K_i$; we sum over these with equal weight 
(normalized to $1$) to define the ``partitioned duality scaling ensemble.''

\subsec{A comment on the metric dependence}

All of the ensembles we discussed depend on an explicit choice of the
K\"ahler metric on configuration space.  This is appropriate as much
of the physics depends on this metric, as does the very problem of
finding vacua in supergravity.

Although we know the K\"ahler metric in a few cases, for example
\IIb\ complex structure moduli space in the weak coupling limit,
and can get interesting results this way, our ability to
compute more general K\"ahler metrics is very limited and probably
will remain so for a long time.

One can try to look for results which do not depend on this choice,
such as the index formula \sgindex\ on a compact configuration space.
So far, it seems hard to come up with interesting examples.

One can try to define ensembles which integrate over K\"ahler
metrics.  An example would be
to take a reference metric $\omega_{K_0}$ (say the weak coupling
metric) and use
$$
\int [dK] e^{-|\omega_K-\omega_{K_0}|^2} ;
$$
appropriate measures and geodesic distances are given in the math
literature.  As it stands, this looks like a rather poorly posed
functional integral in high dimensional quantum gravity, and we would
not recommend it.  It might be interesting to integrate over 
a finite basis of variations to test universality claims.

We suspect that progress on this perplexing point will come more
indirectly.  First, one can read formulas like \sgindex\ backwards,
and argue that if we can get independent information about numbers and
distribution of vacua (e.g. by duality), we can infer properties of
the K\"ahler metric.  Second, one can hope that an ``exact solution''
for the superpotential and so forth will pick out a mathematically
natural configuration space $\CC$ extending the ones we understand
now into the strong coupling regime, and that this space will have
natural candidate metrics.  For now, it seems we must try to work
with what we have.

\newsec{Estimates for the number of Standard Models}

In this section, we attempt to use the ideas we discussed to
``estimate the number of standard models'' in a particular framework.
We are not yet in a position to make controlled estimates, but we will
simply try to apply the various estimates for vacuum counting we
discussed to illustrate the ideas, and to see whether or not there is
an issue of predictivity.  After all, if it were obvious that there
were $10^{10}$ vacua or $10^{1000}$ vacua (with a reasonably uniform
distribution) which qualitatively matched the Standard Model, we would
more or less answer our basic question.  On the other hand, if numbers
$10^{100}$ -- $10^{400}$ came out, we need to consider this question
in more detail.  We also want to see the general shape of the
estimates and which factors might dominate.

We discussed how to precisely define the ``number of vacua'' of
string/M theory satisfying certain qualitative properties such as a
given low energy gauge group, and how to estimate it.  Now if one
chose one of these vacua to focus on, one could go a certain ways in
computing its more detailed properties, but it is clear that at
present our ability to do this is very limited.  Indeed, in the
picture of moduli stabilization we discussed earlier, computing
detailed values of couplings seems inherently difficult.

The main idea we will use to try to go further is to claim that the
ensemble of all vacua of string/M theory of a certain type realize
the uniform ensemble in the space of the remaining couplings.  We
will not give arguments beyond what we gave in sections 3 and 4, but it
should be stressed that this is a testable assumption given any
ability to compute couplings, even a statistical computation of the
sort we discussed earlier.  Furthermore, if we found that an actual
ensemble was not uniform, we would not have to give up -- rather,
we could propose another candidate ensemble which could better model
the true ensemble, and try to draw conclusions from that.

The basic number characterizing our knowledge of the Standard Model
is the volume in coupling space measured in some natural ensemble.
We quoted a number for this in the introduction, but obviously any
number involves many assumptions; let us make this a bit more precise.

The main assumption we made in the estimate we gave in the
introduction was that all Standard Model couplings were distributed
independently and uniformly.  The assumption of independence is of
course false in almost all models of physical interest.  For example,
in grand unified models the three gauge couplings are not independent.
Another class of models might try to explain the structure of the
Yukawa couplings and mass matrices.

The questions of distribution and independence of couplings come to
the fore when one discusses the hierarchy and cosmological constant
problems.  An extreme ``statistical'' point of view would be that
string/M theory produces an ensemble of vacua in which all of the
observed scales and couplings of nature are uniformly distributed.
With enough vacua, one would be likely to well approximate our world.
This scenario would seem rather unpromising for any sort of
predictivity, but as we discussed it has not been ruled out.  Probably
the best hope for ruling it out (assuming there are not too many
models) would be to show that the mass gap (hierarchy) does not come
out with a uniform distribution; if it had (say) a narrow Gaussian
distribution centered on the Planck mass, a modestly large number of
vacua would not be problematic.

A more appealing scenario is one in which the hierarchy is produced by
a mechanism such as the traditional exponentially small
nonperturbative effect in the hidden sector, a large extra dimension
effect or otherwise.  Now if one had a good approximation to the exact
ensemble of string/M theory vacua, the existence of such models would
probably show up as non-analyticity or even a divergence in the
distribution of mass gaps near zero.  This might be an interesting
idea to pursue, but what we will do here is instead simply restrict
attention to the subclass of models which realize supersymmetry breaking
at a hierarchically small scale.  This replaces the strict computation
of a distribution summed over all models, with the computation of the
distribution of this subclass of models.

Within the subclass of models with supersymmetry breaking at $10
\TeV$, and assuming a uniform distribution for the couplings, the
expected probability to realize the Standard Model with an acceptably
small cosmological constant, is $10^{-60-10-9-9-50} \sim 10^{-138}$.
Now it is very likely that the fraction of models which implement
the hierarchy in this way is larger than $10^{-100}$ and that this
is by far the more likely way to realize the physics we observe.
Unfortunately this observation still does not rule out the ``purely
statistical'' scenario, and any systematic discussion must take into
account both possibilities.

Let us proceed to estimate the number of Standard Models coming
from brane constructions which realize the Standard Model gauge
group by wrapping three types of branes on three cycles of distinct
homology class.  As has been noted by Ibanez \ibanez, the most obvious
disadvantage of this class of model is that the gauge couplings do
not naturally unify: the gauge coupling for a brane wrapped on a
$d$-dimensional cycle $\Sigma$ is (up to$2\pi$'s)
$$
{1\over g_{YM}^2} = {{\rm Vol}(\Sigma) \over l_s^d g_s}
$$
and cycles of different homology class have no reason to have
the same volume.  Now the grand unification of the three gauge
couplings seems to be one of the best motivated extrapolations 
we can make beyond observable energies, and this is certainly a 
discouraging observation.  

On the other hand, according to the rules we are playing by here, it
is a problem but only in a particular quantitative way.  If the
couplings had unified, we would treat not all three as independent
variables, leading to a naive estimate like $(1/25)^3$ for the
probability of getting it from the uniform ensemble, but instead 
get a single $1/25$ for the probability of getting the unified coupling.

Under either assumption, the true probability to match the Standard
Model has an additional $10^{-5}$ or so coming from the observational
accuracy of the determination of $\alpha_2$ and $\alpha_3$.  If we can
compute threshold corrections and all other influences on these
couplings very precisely, we can further restrict attention to models
which get this level of structure right.

Now, compared to the other numbers which are entering into our
considerations, these are all relatively small factors.  While the
couplings at the GUT scale are probably the most computable numbers we
can get from string/M theory, they would be expected to depend on
moduli in the general way we discussed before \threshold,\foot{
But note that in some models, these corrections are independent of moduli.
\friedmann}, 
so it is not completely obvious that one
can hope to compute even these uniquely.

Anyways, the point we are trying to make from this discussion is that
for the basic question under discussion, unnaturalness of gauge
coupling unification in a class of models is not a major disadvantage
of the sort that not solving the hierarchy problem would be.
Thus, if we were to find more than $O(10^{138})$ ways to construct
the Standard Model in this framework, we would again face loss of
predictivity.

From our previous discussion, one might expect the main contribution
to the counting to be the number of possible choices of flux.  This is
probably a controllable part of the problem, so this might be good
news, but we should try to check this intuition.

The qualitative features we will assume in our discussion are dynamical
supersymmetry breaking, and the gauge group and chiral matter spectrum
of the Standard Model.  Another feature one might want to include,
which we will not discuss, is the tuning away of dimension 5 operators
required to get acceptable rates of proton decay.  Our excuse for this
will be to say that in a non-unified theory, the natural suppression
of dimension 5 operators would be $1/M_{pl}$, which would suffice.

\subsec{The conditions for Standard Model matter}

The basic structure of brane constructions of the Standard Model has
been given in many works such as \ibanez.
One realizes $SU(2)\times SU(3)$ gauge symmetry
by taking a configuration with two copies of the same brane $B_2$ and
three copies of a different $B_3$.  To avoid adjoint matter, one takes
rigid branes.  One generally needs distinct branes $B_1$ and $B_1'$
associated to ``hypercharge'' to get the usual structure of two Higgs
doublets.  All brane constructions contain many $U(1)$ gauge groups,
most of which are broken or anomalous.  The question of which ones
remain unbroken is somewhat complicated and we will ignore this,
though the need for anomaly cancellation helps in getting a Standard
Model $U(1)$.

Thus, the chance to get the gauge group right on general grounds is
roughly the fraction of brane configurations with 
rigid branes 
$$
B_1 + B_1' + 2 B_2 + 3 B_3 + {\rm other\ branes} .
$$

To get the matter right, one needs branes with particular
intersection numbers: 
\eqn\stdint{\eqalign{
\vev{B_2,B_3} &=\vev{B_3,B_1}=\vev{B_3,B_1'}=\vev{B_1.B_2}=3 ; \cr
\vev{B_1,B_1'} &= \vev{B_1,B_2}=0 .
}}
This includes all charged matter except the right handed electron
(which must appear to get anomaly cancellation) and the Higgs doublets
(which are nonchiral).  
Of course there are more conditions on matter, superpotential and
so forth, which we will ignore here.

To do the problem right, one must work in type \I\ or \IIb\
orientifold theory, and choose an orientifolding.  In the class of
brane constructions we discuss, a large set of orientifoldings just
consist of identifying $\BZ_2$ reflection symmetries of the quiver and
projecting the fields under such a reflection combined with complex
conjugation.  This changes the problem and the estimates we will
discuss but not in a qualitative way (since $\BZ_2$ symmetries are
fairly generic).  Since the main point of the discussion here is not
to get a precise number but rather to illustrate the ideas, we omit
this part of the problem and count brane constructions in type \II.\foot{
Another excuse for this is that doing this right with our present
strategy requires a better understanding of Seiberg duality on the
orientifolded theories.}

The data we just described is computable for each \CY3, but what
we will now assume is that the rigid branes $B_i$ and $B_i'$ are
the basis branes of one of the Seiberg dual theories which can
arise from the \CY.  We must take a set which cancel anomalies;
rather than find this explicitly, we will grant that this can be
done by using $L=O(c_2(M))$ different elementary branes, and use
this as the number of gauge groups in the quiver.
We can then estimate the number of theories
which realize a given intersection form by appealing to our model
ensembles of quiver theories from section 4.  We start with the
duality scaling ensemble.

The number of subsectors that a given quiver theory contains 
which realize \stdint\ is simply the number of ways of selecting
a $4\times 4$ submatrix of $I_{ij}$ which matches the data \stdint.
Given an $L\times L$ matrix, the number of ordered choices will be
$L!/(L-4)! \sim L^4$, and we have six matrix elements to match.  
One of these is actually fixed by $SU(3)$ anomaly cancellation; there
is also an overall factor of $2$ since one can flip all chiralities.
The resulting fraction of models is
\eqn\stdprob{
2 d\mu(0)^2 d\mu(3)^3 = 2 {\alpha^2 \beta^3 \over 3^3} \sim 3 \times 10^{-6} ,
}
so a quiver gauge theory with $L$ nodes  randomly chosen from our
ensemble will typically realize the Standard Model in $3\times 10^{-6} L^4$
different ways.

This is typically a large number, but a much more stringent condition
is that the resulting candidate Standard Model does not contain
exotic matter charged under the Standard Model gauge group.  Besides the
fact that it has not been seen, the main problem with this is that
it spoils grand unification, a constraint we are not imposing, but
let us anyways estimate the probability to not have such matter.
If we only worry about $SU(2)\times SU(3)$, this will be
$$
d\mu(0)^{2(L-4)} \sim 4^{-(L-4)}
$$
which is a major suppression.  Indeed, in explicit brane constructions, it
tends to be difficult to eliminate such exotic matter.\cvetic

We can now sum the resulting estimate over our list of \CY3's.  The
number $L=c_2(M)$ can be computed for each, but we will just take
$L \sim K \sim \chi$.  This
gives the estimated number of Standard Model quivers
\eqn\smquiv{
N_{SMQ\ d.s.} = 3\times 10^{-6} \times \sum_{K=0}^{400} K^5 4^{-K}
\sim 10^{-2} .
}

If we trusted our model ensemble, the fact that this estimate
comes out less than $1$ would mean that the Standard Model was in fact
difficult to realize, because of the difficulty of eliminating exotic
charged matter.  Of course the approximate nature of the estimate
means that solutions could well exist, but probably at low $\chi$;
direct search through the low $\chi$ Calabi-Yau's would be quite
interesting in this case.

As we discussed earlier, the assumption of complete independence made
in the duality scaling
model ensemble is probably false, because one can have
groups of cycles which do not intersect with other groups.  The
constraint of no exotic matter would thus favor realizing the Standard
Model in such a group, say with branes wrapped on cycles obtained by
resolving one isolated singularity.  This observation has been made in
the brane construction literature, but without quantitative
considerations, it is hard to know how much significance to give this
constraint compared to other constraints one might try to realize.

A similar estimate for the partitioned 
duality scaling ensemble can be obtained
by using the fraction of partitions of $K$ which contain an integer $M$,
which goes as $\log K/M$, to obtain
\eqn\psmquiv{
N_{SMQ\ partitioned} = 3\times 10^{-6} \times \sum_{K=0}^{400} K \log K
\sum_M M^3 4^{-M} 
\sim 75 .
}
This would certainly be an interesting estimate if true, as it
suggests a sparse set of solutions scattered among the various
\CY3's.  At this point we are not claiming it is reliable.  Rather,
our point is that one can do a much better job of characterizing the
true ensemble of gauge theories realizable by brane constructions with
existing techniques, and a reliable estimate would be of great value
in deciding how to search through the large set of possibilities.

To continue, let us grant
\psmquiv\ as a factor in the total estimate, which is
$$
N_{SM} = 75 \times N_{vac}({\rm non\ SM\ branes}) .
$$
The second factor is of the type which we discussed in section 3
and gave the generic estimate $c^N$ for some $c>1$.  Thus it is
expected to be large, but smaller than
the number of vacua which would be obtained by counting all
combinations of branes, not separating out some to realize the
Standard Model.

\subsec{The conditions for low energy supersymmetry breaking}

A straightforward way to approach this question is to ask for a gauge
sector which dynamically breaks supersymmetry.  The status of this
problem is reviewed in \ss.  At present there is no general
classification of such models, but there are special cases which are
well understood such as the $(3,2)$ model.  This is quite
similar to the Standard Model but with intersection numbers $1$
instead of $3$.  Thus we need more branes $B_1'' + 2 B_2' + 3 B_3'$
with specified intersection numbers, and get an estimate $d\mu(1)^3
\sim 10^{-4}$.  The main difference from our previous estimate
is that we will not worry about exotic matter.  While exotic matter
brings in the real possibility of additional flat directions which
spoil supersymmetry breaking, it is also known that there are many
more supersymmetry breaking theories, so some possibilities may work.

Another supersymmetry breaking mechanism more in tune with the ideas
developed here is simply to observe that a flux superpotential would
be expected to contain many supersymmetry breaking minima, simply on
grounds of genericity.  Indeed, the explicit gauge models may well be
dual to this type of realization.  An advantage of this point of view
is that much of the issue in finding supersymmetry breaking is in
showing that the moduli are stabilized in a reasonable regime after
supersymmetry breaking, so one needs an approach in which this can be
done.  This type of question can be studied in the simplified
ensembles of section 4, as we will discuss in \refs{\dougashok,\dougzel}.

At the present state of knowledge, it is difficult to do better on
this problem, and we will guess that a fraction $10^{-4}$ of
models contain dynamical supersymmetry breaking.  One also needs to
estimate the probability that the supersymmetry breaking scale comes
out right (not hard if we grant the usual exponential suppression) and
that this sector is coupled to the observable sector in an acceptable
way (e.g. which solves the $\mu$ problem); we will also assume this is
not hard, say that $O(10^{-3})$ of the models do it.  

While these particular numbers have no real significance, the basic
assumption we are making is that since we do not (yet) observe the
supersymmetry breaking sector, one could pass this test in many ways.
Using this estimate, one obtains
$$
N_{SM + SUSYB} = 10^{-5}
 \times N_{vac}({\rm non\ SM\ and\ SUSY\ branes}) .
$$

\subsec{The number of models}

We now have an estimate which factors into the number of ways
to realize the structure assumed in our class of models, and
the number of vacua which correspond to choices in the hidden sector.
We gave various estimates for the latter, based on considering the
hidden sector as made up of branes or made up of fluxes.  

A simple minded way to get to a final result is to say that since our
construction separated out $12$ branes as special, we have $N-12$
branes in the hidden sector, and a vacuum multiplicity of the
order $c^{N-12}$.  This would be multiplied by the number of choices
of flux.

However, it is clearly not the case that these choices are
independent; for example the geometric dualities of Gopakumar and Vafa
\gv\ are an obvious redundancy, and there are surely many more.
This is the largest ``systematic uncertainty'' in our counting but a
point on which theoretical progress can be made.  

For present purposes, let us assume that all ``non Standard Model''
and ``non susy breaking'' branes can be dualized to flux.

This is an interesting assumption because after dualizing, we are
left with a very specific
configuration of $12$ Dirichlet branes of $6$ distinct types.  This is
typically not what will come out of anomaly cancellation for \CY3's
with large $K=b_{1,1}$; we might instead expect $O(K)$.  Thus many of
the \CY3's need not be considered at all under this assumption, as all
of these brane configurations are actually redundant.

The number of \CY3's with $b_{1,1}\sim 6$ is much smaller, say
$400-1000$.  We could then directly apply \stdprob\ and our ``estimate''
for the likelihood of successful supersymmetry breaking to obtain
\eqn\finalSM{\eqalign{
N_{SM} &\sim 3 \times 10^{-13} \times \sum_{CY_3\ {\rm with}\ K\sim 6} 
 N_{flux\ vac} \cr
 &\sim 10^{-10} \times \vev{ N_{flux\ vac} }
}}
in terms of the average number of flux vacua in this class (a sum which
is probably dominated by the \CY3's with large $b_{2,1}\sim 400$).

If we trusted this number, the upshot would be that realizing the
Standard Model with low energy supersymmetry breaking is not easy,
but not so difficult considering the expected number of flux vacua.
While the number of qualitatively correct models would be large, the
question of how many matched the couplings would depend on the number
of flux vacua and the resulting distribution of couplings.

We are not going to defend this number very strongly.  But we will
defend the discussion which led up to it, as illustrating a new and
different way to think about ``string phenomenology,'' and suggesting
all sorts of new questions about both the elements of string
compactification and the types of theories which might lead to
observable physics, which will be interesting to explore.

We believe that reliable estimates of this type could be made without
having exact results for string/M theory at all couplings, but simply
with better theoretical understanding of some key points which emerged
in our discussion, and a good deal of work.  A consistency check which
one could apply to the results would be to make estimates for a
variety of dual realizations of the same family of vacua, and see if
one gets rough agreement.

Not having a reliable estimate, we would still conjecture that the
qualitative structure of the Standard Model is the result of discrete
choices which are not that hard to realize, and that the 
fraction of models which meet the qualitative tests (ignoring values of
couplings) is closer to $O(10^{-10})$ than to $O(10^{-100})$.

There is independent evidence for this, in that among the few models
which have been considered in any depth, one does get Standard Model
candidates or at least ``near-misses'' (say with exotic charged
matter).  One might worry that this is a selection effect ({\it i.e.},
people only study models which are likely to realize the Standard
Model), but this is clearly not true, as almost none of the \CY3's on
the list of \kreusch\ have been considered at all.

If this estimate is even approximately valid, this shows that the main
problem is to get a good estimate for numbers of flux vacua.  Although
an estimate of $O(10^{100})$ seems plausible, it is not at all ruled
out that there are \CY3's or brane gauge theories with extremely large
numbers of vacua, so the question of testability remains open.

\newsec{Conclusions}

In this work, we proposed a new approach to the ``vacuum selection
problem'' of string/M theory.  We believe that we should not postpone
work on this problem until either an ``exact solution'' or some key
``Vacuum Selection Principle'' (or both) are discovered.  Rather, we
should learn to work better with the many known ``vacuum selection
principles'' (more simply, ``tests'') of fitting observation and other
well motivated theoretical frameworks such as cosmology, by getting a
rough overall picture of the set of all string/M theory vacua and
estimates of how difficult it is to satisfy each of the various tests,
meaning what fraction of the total number of vacua pass a test or
combination of tests.  We believe such estimates will be invaluable
for any systematic program to test string/M theory, even as better
selection principles emerge.

Making such estimates requires working with large numbers of models in
a uniform way, which is only practical if one has a systematic
construction.  We discussed \IIb\ brane/flux compactification on \CY3,
which we believe is approaching the level of sophistication needed.
We identified several theoretical points which need to be clarified to
make a proper discussion, such as the scope of geometric dualities in
compactifications with both branes and flux.

We then proposed to simplify and perhaps make progress on this
challenging problem by looking for simpler ensembles of effective
theories which well approximate the ensemble of effective theories
which actually comes out of string/M theory, in the sense that the
correct estimated fraction of vacua passing a particular test is
reproduced by the ensemble.  It is worth emphasizing that this project
has meaning if and only if a ``true ensemble of effective theories coming
from string/M theory'' exists.  While this is not proven, it only
depends on two essential claims -- namely, that string/M theory has a
precise definition in which the candidate physical vacua can be
described by low energy effective theories, and that the set of these
effective theories is ``locally finite'' (as we discussed), possibly
after imposing physical restrictions such as those we discussed ({\it
e.g.}  the bound on compactification volume).  The first claim is
supported by the fact that physical observation can be fit with
effective field theory, while the failure of the second would probably
eliminate any sort of predictivity, and both claims are consistent
with our present understanding of string/M theory, so we consider them
as very reasonable hypotheses on which to base further considerations.

Granting the existence of this ensemble, the study of candidate
approximations to it is a well posed problem, and well posed questions
in string/M theory tend to have simple answers.  To begin this study,
we gave examples of simple ensembles, which have some rough similarity
to the real problems coming out of string/M theory, and illustrations
of the type of computations one could do with them.  The basic
computation one can do is to find expected numbers of supersymmetric
and nonsupersymmetric vacua, and the dependence of these numbers on
parameters of the ensemble.  We also illustrated how dualities can be
used to constrain these ensembles.

We then made some rough estimates of numbers of Standard Models which
suggest that in \IIb\ theory, the main ingredient in getting a good
estimate is to estimate the number of possible choices of flux.  There
are many other ingredients which need to be refined, among which the
problem of whether and in what sense a ``uniform'' ensemble of
effective theories will emerge.

We stress that, although the specific constraints of realizing the
Standard Model, supersymmetry breaking and so forth, have of course
been much studied, before this work there has been no way to quantify
how constraining each of these considerations might be within string
theory.  By considering ensembles, this can be quantified, giving us a
framework in which to systematize, evaluate and combine these
considerations.

Although the ensembles we considered are somewhat crude, we can
progress by formulating better ones which try to reflect more of the
structure of the problem, and test our hypothesized ensembles against
statistics of sample sets of string vacua constructed either
systematically, or by choosing random examples and doing detailed
model-by-model analysis.  By finding better ensembles, we will be
improving our understanding of the distribution of string/M theory
vacua in a relatively concrete way.  One might think of the structure
of a good ensemble as capturing a ``stringy'' concept of naturalness,
which could improve on traditional ideas of naturalness in guiding
string phenomenology.

Suppose we had these estimates: what would we do next?  The best
argument that they are worth having is that what we would do next
depends very much on what comes out.  Obviously one would want to
focus on tests which seem difficult to meet yet are theoretically
tractable; it is not {\it a priori} obvious which ones these are.  

We even argued that depending on what comes out, we might find that
string/M theory has much less predictive power than we thought,
perhaps none.  At present it is reasonable to think that string/M
theory will have predictive power, but we should admit that we do not
really know, and try to find out.

\penalty-100
\medskip 
Acknowledgements

This work originated in the talk \jhstalk, and has benefitted from
discussions with many people, particularly Bobby Acharya, Sujay Ashok,
Paul Aspinwall, Tom Banks, Rafael Bousso, Philip Candelas, Mirjam
Cvetic, Frederik Denef, Savas Dimopoulos, Tamar Friedmann, Steve
Giddings, Mark Gross, Shamit Kachru, David Kutasov, Sheldon Katz,
Alistair King, Maxim Kontsevich, Hong Liu, Juan Maldacena, Greg Moore,
Lubos Motl, Miles Reid, Paul Seidel, Savdeep Sethi, Steve Shenker,
Bernie Shiffman, Nati Seiberg, Richard Thomas, Sandip Trivedi, Cumrun
Vafa, Edward Witten, Steve Zelditch and Chen-Gang Zhou.  I would also
like to acknowledge general inspiration drawn from the works of
Mikhail Gromov.  Finally, I thank the Banff International Research
Station for providing a pleasant environment for the completion of
this work.

This research was supported in part by DOE grant DE-FG02-96ER40959.

\listrefs

\end